\documentclass[pra,preprint,aps,showpacs]{revtex4-1}
\usepackage{graphicx}

\newcommand{\om}{\omega}

\newcommand{\E}{{\bf E}}
\newcommand{\ev}{{\bf e}}

\newcommand{\rv}{{\bf r}}

\newcommand{\Hv}{{\bf H}}

\newcommand{\pv}{{\bf p}}

\newcommand{\uv}{{\bf u}}

\newcommand{\shat}{\hat{s}}
\newcommand{\ppp}{\hat{p}^+}
\newcommand{\ppm}{\hat{p}^-}

\newcommand{\green}{{\stackrel{\leftrightarrow}{\bf G}}}
\newcommand{\iden}{{\stackrel{\leftrightarrow}{\bf I}}}

\newcommand{\gamperp}{{\stackrel{\leftrightarrow}{\bf \Gamma}}}
\newcommand{\deter}{{\stackrel{\leftrightarrow}{\bf D}}}
\newcommand{\dyadic}{{\stackrel{\leftrightarrow}{\bf A}}}
\newcommand{\dyadicb}{{\stackrel{\leftrightarrow}{\bf B}}}
\newcommand{\dyadicc}{{\stackrel{\leftrightarrow}{\bf C}}}

\newcommand{\alpht}{{\stackrel{\leftrightarrow}{\bf \alpha}}}
\newcommand{\bett}{{\stackrel{\leftrightarrow}{\bf \beta}}}
\newcommand{\mv}{{\bf m}}

\begin{document}

\title{Strong tip-sample coupling in thermal radiation scanning tunneling
microscopy}
%
\author{Karl Joulain}
\affiliation{Institut PPrime, CNRS-Universit\'e de Poitiers-ENSMA, UPR 3346, 2, Rue Pierre Brousse, B.P 633, 86022 Poitiers Cedex, France.}
\author{P. Ben-Abdallah}
\affiliation{Laboratoire Charles Fabry, UMR 8501, Institut d'Optique, CNRS, Universit\'e Paris-Sud 11, 2 avenue Augustin Fresnel, 91127 Palaiseau Cedex, France}
\author{P.-O. Chapuis}
\affiliation{Centre de Thermique de Lyon (CETHIL), 
CNRS, INSA de Lyon, UCBL, Campus de la Doua, 69621 Villeurbanne, France}
\author{Y. De Wilde}
\affiliation{Institut Langevin, ESPCI Paris Tech, CNRS, 1 rue Jussieu, 75005 Paris, France}
\author{A. Babuty}
\affiliation{Institut Langevin, ESPCI Paris Tech, CNRS, 1 rue Jussieu, 75005 Paris, France}

\author{C. Henkel}
\affiliation{Institute of Physics and Astronomy, University of Potsdam, 
Karl-Liebknecht-Str. 24/25, 14476 Potsdam, Germany}

\date{Thu 07 Nov 2013}

\begin{abstract}
We analyze how a probing particle modifies the infrared electromagnetic near field
of a sample. The particle, described by electric and magnetic polarizabilities, represents the tip of an apertureless scanning optical near-field microscope (SNOM). We show that the interaction with the sample can be accounted for by ascribing to the particle dressed polarizabilities that combine the effects of image dipoles with retardation. When calculated from these polarizabilities, the SNOM signal depends only on the fields without the perturbing tip. If the studied surface is not illuminated by an external source but heated instead, the signal is closely related to the projected electromagnetic local density of states (EM-LDOS). 
Our calculations provide the link between the measured far-field spectra and the 
sample's optical properties.
We also analyze the case where the probing particle is hotter than the sample and evaluate the impact of the dressed polarizabilities on near-field radiative heat transfer. We show that such a heated probe above a surface performs a surface spectroscopy, in the sense that the spectrum of the heat current is closely related to the local electromagnetic density of states.
The calculations agree well with available experimental data.

\end{abstract}

\maketitle

\section{Introduction}
Since the seminal work of Rytov and co-workers \cite{Rytov:1989ur}, it is known that thermal radiation has a different behaviour when the involved characteristic lengths are large or small compared to the thermal wavelength \cite{Joulain:2005ih,Volokitin:2007el,Dorofeyev:2011bg}. For example,
the heat flux transferred between bodies separated by
a subwavelength distance can exceed by far the one between black bodies \cite{Polder:1971uu,BenAbdallah:2010hp}. Energy density \cite{Shchegrov:2000td} and coherence properties \cite{Henkel:2000tr} are also strongly affected in the near field, especially close to materials exhibiting resonances such as polaritons. Knowing precisely how the electromagnetic field behaves close to a surface is therefore an important issue in order to address potential applications involving near-field heat transfer.

From an experimental point of view, the coherence properties of near-field radiation have been utilized to produce directional and monochromatic thermal sources \cite{Greffet:2002ur,Lee:2006cj,Biener:2008cj}. 
The enhancement of radiative heat transfer at short distances has been 
demonstrated recently between two macroscopic surfaces \cite{Ottens:2011kh,Kralik:2012},
but probe microscopy techniques are still playing a prominent role
\cite{Kittel:2005fr,Narayanaswamy:2008gj,Rousseau:2009es}. Near-field thermal flux imaging has been operated with a scanning thermal microscope \cite{Kittel:2008bc,Wischnath:2008hp}. A scanning near-field optical microscope (SNOM) without external illumination, termed thermal radiation scanning tunneling microscope (TRSTM), has also been used to image surfaces \cite{DeWilde:2006kt,Kajihara:2010fo,Kajihara:2011uu}. Very recently, local spectra have also been measured \cite{Babuty:dYmzb9en,Jones:2012fx}.

Most of these experimental techniques use a small probe brought in the vicinity of the 
sample surface. 
%
Its response to the sample's
near field is given by a polarizability. The induced multipoles are sources that radiate into the far field, thus providing the TRSTM signal. 
At short (sub-wavelength) distances however, the mutual interaction between the probe 
and the surface modifies the local electromagnetic field~\cite{GarciadeAbajo:2007eb,Intravaia:2010gp,BenAbdallah:2011be,Castanie:2011ue}, 
and this actually changes the probe's optical properties such as the polarizability.
These interactions also complicate the data analysis for the near-field techniques
mentioned above. In particular, one is often interested in the sample's optical 
properties, as encoded in the electromagnetic local density of states (EM-LDOS)~%
\cite{Joulain:2003hc,Kittel:2008bc}. Due to the tip-sample interaction, it is no
longer obvious how the TRSTM signal scattered by the tip into the far field is 
related to the EM-LDOS. In particular, can a SNOM detecting thermal radiation be 
the electromagnetic equivalent of the scanning tunneling microscope detecting 
the electronic LDOS \cite{Tersoff:1985wm}? Moreover, one can ask 
what information
can be extracted from the exchanged heat flux between the probe and the sample.

If some of these questions have already been addressed in the past~\cite{Joulain:2003hc,Mulet:2001kp}, our goal is here to clarify remaining interrogations.  Following previous similar works~\cite{Knoll:2000wm,Sun:2007cl}, 
we will first see how the particle polarizability 
can be replaced by an effective or {\it dressed} polarizability taking into account 
multiple reflections between the probe and the surface. %
This is more general than the image-dipole model~\cite{Knoll:2000wm} 
whose range of validity is very restricted in the infrared.
We will then use the theory to calculate the signal detected in the far field 
when the near field is scattered by a probe dipole. 
An expression for the SNOM signal is calculated and illustrated 
by scanning a surface excited either by a plasmon or by 
broadband thermal radiation (TRSTM mode).
In this paper, the probe tip is modeled by both electric and magnetic dipoles. 
This approach fails to capture field inhogeneities across the tip that
occur at short distances (comparable to the tip size) and excite higher
multipoles. It has the advantage, however, of providing relatively simple 
expressions that can be physically interpreted. The model has also been shown to reproduce the main physical ingredients in the case of a TRSTM tip \cite{Babuty:dYmzb9en}. It is powerful since the analysis of the results based on analytical expressions is straightforward. It is sure that a more accurate modeling of the tip, e.g. with a cone, would be better. However, the analysis of the various detected components (polarization, electric vs magnetic, etc) may be much more complicated in this case. The dipole approach could 
also be included as a building block into more flexible numerical schemes
like the coupled dipole method~\cite{Keller:1993,Lax:1951tb,GarciadeAbajo:2007eb}, 
the multiple multipole method~\cite{Hafner:1999} or the discrete dipole 
approximation \cite{Yurkin:2007,Loke:2011} that has been very recently 
adapted to thermal near field radiation \cite{Edalatpour:2013ve}. 
It is also an alternative to more complicated but exact numerical methods 
such as surface-integral methods~\cite{Rodriguez:2011ki}.

We conclude the paper by analyzing the signal detected in far field 
due to a heated probe 
when accounting for probe-surface interactions.
Finally, the radiative cooling of a particle in the near field 
and the spectrum of the heat flux are analyzed.

\section{Dressed polarizabilities}
\label{s:dressed-polarizabilities}

\begin{figure}[h]
\centering
\includegraphics[width=10cm]{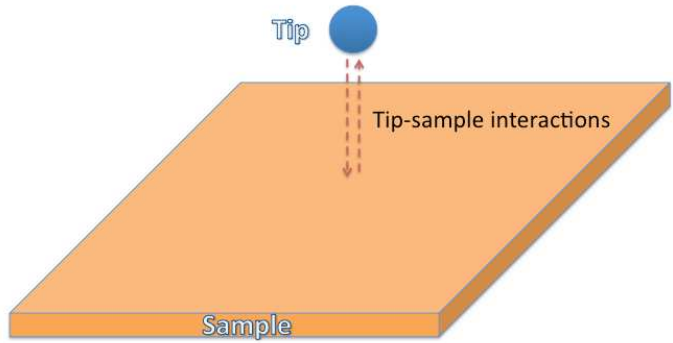}
\caption{Sketch of the system.}
\label{systeme}
\end{figure}

We propose here to calculate the {\it dressed} polarizability of a dipolar particle when it is placed in an environment which is different from free space. Indeed, when a particle is added to a system, the electromagnetic field present in the system illuminates the particle and induces a dipole moment (Fig.~\ref{systeme}). This dipole radiates a field everywhere that scatters back to the particle position. This interaction between the particle and the system modifies the total electromagnetic field, which is no longer given by the field in absence of the perturbing probe. In other words, the probe is no longer a passive test dipole. Our aim is to show that we can work with the unperturbed electromagnetic field if we ascribe to the particle a dressed polarizability.

Let us call $\E^0$ the electromagnetic field in the system without the particle
(i.e., the ``non-perturbed field''). 
When a particle (tip) is placed in the system at position $\rv_t$, an electric dipole $\pv$ and a magnetic dipole $\mv$ will be induced in the particle (tip). These dipoles radiate a field: the total field $\E^{tot}$ is the sum of $\E^0$
and of the field radiated by the dipoles. In the following, we use Green tensors to 
express the field radiated by a dipole:
\begin{equation}
\label{ }
\E^{tot}(\rv) = \E^{0}(\rv) + \green^{EE}(\rv,\rv_t)\cdot\pv 
	+ \green^{EH}(\rv,\rv_t)\cdot\mv
\end{equation}
where $\green^{EE}$ and $\green^{EH}$ take into account the reflection (scattering)
by the sample. 
This can also be written
\begin{equation}
\label{eloc}
\E^{tot}(\rv)=\E^{0}(\rv)+\green^{EE}(\rv,\rv_t)\cdot\alpha\E^{tot}(\rv_t)+\green^{EH}(\rv,\rv_t)\cdot\beta\Hv^{tot}(\rv_t)
\end{equation}
where $\alpha$ and $\beta$ are the ``bare'' electric and magnetic polarizabilities 
of the particle: they 
do not know about the surrounding sample
and describe its reaction to the local field $\E^{tot}(\rv_t)$, $\Hv^{tot}(\rv_t)$.
For a spherical particle, they are scalars. The generalization to anisotropic 
particles where $\alpha$ and $\beta$ become tensors is straightforward, see,
e.g., Ref.\cite{Huth:2010fg}.

Analogous expressions exist for the magnetic field
\begin{equation}
\label{}
\Hv^{tot}(\rv)=\Hv^{0}(\rv)+\green^{HE}(\rv,\rv_t)\cdot\pv+\green^{HH}(\rv,\rv_t)\cdot\mv
\end{equation}
that can also be written
\begin{equation}
\label{hloc}
\Hv^{tot}(\rv)=\Hv^{0}(\rv)+\green^{HE}(\rv,\rv_t)\cdot\alpha\E_{tot}(\rv_t)+\green^{HH}(\rv,\rv_t)\cdot\beta\Hv^{tot}(\rv_t)
\end{equation}
The Green tensors used here come in four types: $\green^{EE}(\rv,\rv_t)$ gives the electric field at position $\rv$ when an electric dipole source is placed at $\rv_t$. In the same way, $\green^{EH}(\rv,\rv_t)$ gives the electric field at position $\rv$ when a magnetic dipole is placed at $\rv_t$.  $\green^{HE}$ and $\green^{HH}$ respectively give the magnetic field of an electric and a magnetic dipole. 
These Green tensors, which can all be calculated from 
$\green^{EE}(\rv,\rv_t)$ \cite{Joulain:2010bq}, are the sum of a direct 
contribution (i.e., the Green tensor \emph{in vacuo}) and of a contribution 
due to scattering from the sample. 
The latter is labelled in the rest of the paper by a subscript $R$.
When one considers the electromagnetic field at the particle position $\rv_t$, 
the direct contribution leads to renormalized parameters (radiative line width, 
Lamb shift) that we suppose already included in $\alpha$, $\beta$, 
so that we may focus on the scattered Green tensors only. 

By solving the system (\ref{eloc}, \ref{hloc}), we find the local field in the form
\begin{eqnarray}
\E^{tot}(\rv_t) & = &\dyadic\ \E^0(\rv_t)+\dyadicb \ \Hv^0(\rv_t) \label{elocfull} \\
\Hv^{tot}(\rv_t) & = & \dyadicc\ \E^0(\rv_t)+\deter\ \Hv^0(\rv_t)\label{hlocfull} 
\end{eqnarray}
where 
\begin{eqnarray}
	\label{eq:A-tensor}
\dyadic & = & \left[[\iden-\alpha\green^{EE}_R(\rv_t,\rv_t)]-\alpha\beta\green^{EH}_R(\rv_t,\rv_t)[\iden-\beta\green^{HH}_R(\rv_t,\rv_t)]^{-1}\green^{HE}_R(\rv_t,\rv_t)\right]^{-1} \\
\dyadicb & = &\left[-\alpha\green^{HE}_R(\rv_t,\rv_t)+\beta^{-1}[\iden-\beta\green^{HH}_R(\rv_t,\rv_t)][\green^{EH}_R(\rv_t,\rv_t)]^{-1}[\iden-\alpha\green^{EE}_R(\rv_t,\rv_t)]\right]^{-1}\\
 \dyadicc & = & \left[-\beta\green^{EH}_R(\rv_t,\rv_t)+\alpha^{-1}[\iden-\alpha\green^{EE}_R(\rv_t,\rv_t)][\green^{HE}_R(\rv_t,\rv_t)]^{-1}[\iden-\beta\green^{HH}_R(\rv_t,\rv_t)]\right]^{-1}\\
 \deter& = & \left[[\iden-\beta\green^{HH}_R(\rv_t,\rv_t)]-\alpha\beta\green^{HE}_R(\rv_t,\rv_t)[\iden-\green^{EE}_R(\rv_t,\rv_t)]^{-1}\green^{EH}_R(\rv_t,\rv_t)\right]^{-1}
	\label{eq:D-tensor}
\end{eqnarray}
The preceding equations are fully general for the total field at the dipole position. 
Apart from the fact that the magnetic dipole is taken into account, the reasoning used 
here to obtain the total field (also known as ``self-consistent field'')
is very similar to previous works using the so-called coupled dipole theory, see, 
e.g.~\cite{Keller:1993,Lax:1951tb,GarciadeAbajo:2007eb,Intravaia:2010gp,BenAbdallah:2011be}. 
Using the bare polarizabilities, the induced dipoles can be related to the non-perturbed 
fields at the tip position 
\begin{equation}
\label{eq:def-dressed-alpha}
\left(\begin{array}{c}
      \pv   \\
      \mv
\end{array}\right)=\left(\begin{array}{cc}
    \alpha\dyadic  & \alpha\dyadicb    \\
    \beta\dyadicc  &   \beta\deter
\end{array}\right)\left(\begin{array}{c}
      \E^0( \rv_t )  \\
      \Hv^0( \rv_t )
\end{array}\right)
\end{equation}
The four sub-matrices can be seen as dressed polarizabilities
that depend on the particle position $\rv_t$;
they will be anisotropic in general.

We now highlight the case of a simple system made of a planar interface separating a material and vacuum. The corresponding Green tensors are well known \cite{Sipe:1987td,Joulain:2010bq} and depend on the material's optical properties. 
In this plane-parallel geometry with $\hat{\bf z} = (0, 0, 1)^T$ normal to the
interface, it 
is convenient to bring the tensors into block-diagonal form by identifying 
suitable sub-spaces, in particular to perform the inversions in 
Eqs.(\ref{eq:A-tensor}--\ref{eq:D-tensor}).
The resulting forms are 
\begin{equation}
\label{ }
\green^{HE}_R(\rv_t,\rv_t)=\mu_0\green^{EH}_R(\rv_t,\rv_t)=\left(\begin{array}{ccc} 0 & a & 0 \\ -a  & 0 & 0 \\0 & 0 & 0\end{array}\right)
\end{equation}
\begin{equation}
\label{eq:ee-and-hh-Green}
\green^{EE}_R(\rv_t,\rv_t)=\left(\begin{array}{ccc} b & 0 & 0 \\ 0  & b & 0 \\0 & 0 & C\end{array}\right), 
\qquad
\green^{HH}_R(\rv_t,\rv_t)=\left(\begin{array}{ccc} d & 0 & 0 \\ 0  & d & 0 \\0 & 0 & f\end{array}\right)
\end{equation}
with matrix elements
\begin{eqnarray}
a & = & \frac{\omega}{8\pi} \int_0^\infty KdK(r^s-r^p)e^{2i\gamma z_t} 
\label{eq:HExy-Green}
\\
b & = & \frac{i\mu_0\omega^2}{8\pi}\int_0^\infty\frac{KdK}{\gamma}(r^s-r^p\frac{\gamma^2}{k_0^2})e^{2i\gamma z_t} \\
C & = & \frac{i\mu_0\omega^2}{4\pi}\int_0^\infty\frac{K^3dK}{\gamma k_0^2}r^pe^{2i\gamma z_t}\\
d & = & \frac{i\omega^2}{8\pi c^2}\int_0^\infty\frac{KdK}{\gamma}(r^p-r^s\frac{\gamma^2}{k_0^2})e^{2i\gamma z_t}
\label{eq:HHxx-Green}\\
f & = & \frac{i\omega^2}{4\pi c^2}\int_0^\infty\frac{K^3dK}{\gamma k_0^2}r^se^{2i\gamma z_t}
\label{eq:HHzz-Green}
\end{eqnarray}
In these integrals over plane waves, rotational symmetry in the $xy$-plane has 
been exploited: the wave vector projected onto the surface has length $K$, its
perpendicular component is $\gamma=\sqrt{k_0^2-K^2}$ and $k_0 = \omega / c$
is the wavenumber \emph{in vacuo}. $r^s$ and $r^p$ denote the Fresnel reflection amplitudes 
at the vacuum-material interface in the principal polarizations
s (or TE) and p (TM).

The expression~(\ref{eq:def-dressed-alpha}) for the electric and magnetic
dipoles becomes in terms of the dressed polarizabilities
%
\begin{eqnarray}
\pv(\rv_t) & = & \alpht^{EE}\E^0(\rv_t)+\alpht^{EH}\Hv^0(\rv_t)  \\
\mv(\rv_t)& = & \bett^{HE}\E^0(\rv_t) + \bett^{HH}\Hv^0(\rv_t)
\end{eqnarray}
and the latter read as follows in the our planar setting:
\begin{eqnarray}
\label{alpheeff}
\alpht^{EE} &=& \left(\begin{array}{ccc} \frac{\alpha}{1-\alpha b+\frac{\mu_0\alpha\beta a^2}{(1-\beta d)}} & 0 & 0 \\ 0  & \frac{\alpha}{1-\alpha b+\frac{\mu_0\alpha\beta a^2}{(1-\beta d)}} & 0 \\0 & 0 & \frac{\alpha}{1-\alpha C}\end{array}\right)
\\
\label{}
\alpht^{EH} &=& \left(\begin{array}{ccc} 0 & \frac{\mu_0\alpha\beta a}{(1-\alpha b)(1-\beta d)-\mu_0\alpha\beta a^2} & 0 \\ -\frac{\mu_0\alpha\beta a}{(1-\alpha b)(1-\beta d)-\mu_0\alpha\beta a^2}   & 0& 0 \\0 & 0 & 0\end{array}\right)
=
\mu_0 \bett^{HE},
\\
\label{betaeff}
\bett^{HH} &=&\left(\begin{array}{ccc} \frac{\beta}{1-\beta d+\frac{\mu_0\alpha\beta a^2}{(1-\alpha b)}} & 0 & 0 \\ 0  & \frac{\beta}{1-\beta d+\frac{\mu_0\alpha\beta a^2}{(1-\alpha b)}} & 0 \\0 & 0 & \frac{\beta}{1-\beta f}\end{array}\right).
\end{eqnarray} 
%
Due to multiple reflections between particle and interface, these
polarizabilities are anisotropic tensors. Note the dimensionless parameters
$\alpha b$ and $\alpha C$ for the electric case,
and $\beta d$ and $\beta f$ for the magnetic one. We shall see below that
an upper limit to these parameters scales, in order of magnitude, as $(R_t/z_t)^3$ 
where $R_t$ is the radius of the probe particle. The dipole approximation is valid
when the distance $z_t$ is sufficiently large compared to the radius. Otherwise the
image field would be significantly inhomogeneous across the particle volume,
inducing quadrupole and higher multipoles. We therefore
restrict to $(R_t/z_t)^3 \le 1/8 \ll 1$, so that the anisotropy of
$\alpht^{EE}$ is weak. The magneto-electric cross-polarizabilities
$\alpht^{EH}$, $\bett^{HE}$ scale with the parameter $\mu_0 \alpha \beta a^2$ 
which is typically small for the same reason. We see below that we can safely 
neglect them.
In the plots of the following Section, we compare the different polarizabilities after
normalizing them by the volume $V = 4\pi R_t^3/3$ of the particle, namely
$\alpht^{EE} / (\epsilon_0 V)$,
$c \alpht^{EH} / V$,
$\mu_0 c \bett^{HE} / V$,
and $\bett^{HH} / V$.


\section{Parametric study of the dressed polarizabilities}

In this section, we analyze how the dressed polarizabilities depend on the material's optical properties, particle sizes and particle-surface distance. 
For a spherical particle, the bare polarizabilities are well established in Mie theory.
They involve the ratio of the particle radius $R_t$ to the wavelength $\lambda$ 
(Mie parameter $x = 2\pi R_t / \lambda = k_0 R_t$) and the corresponding ratio 
$y = x \sqrt{\epsilon( \omega )}$ inside the particle, whose dielectric function is 
$\epsilon( \omega )$. 
For a non-magnetic material ($\mu=1$), the following expressions given by Chapuis \& 
al.~\cite{Chapuis:2008kcb} apply if the wavelength is much larger than $R_t$: 
\begin{equation}
\label{eq:Mie-alpha}
\alpha(\omega)=\epsilon_02\pi R_t^3\frac{2\left[\sin(y)-y\cos(y)\right]-x^2\left[\frac{-\sin(y)}{y^2}+\frac{\cos(y)}{y}+\sin(y)\right]}{\left[\sin(y)-y\cos(y)\right]+x^2\left[\frac{-\sin(y)}{y^2}+\frac{\cos(y)}{y}+\sin(y)\right]}
\end{equation}
and 
\begin{equation}
\label{eq:Mie-beta}
\beta(\omega)=-2\pi R_t^3\left[\left(1-\frac{x^2}{10}\right)+\left(-\frac{3}{y^2}+\frac{3}{y}\cot (y)\right)\left(1-\frac{x^2}{6}\right)\right]
\end{equation}
For metallic particles or near a resonance of $\epsilon( \omega )$, 
the parameter $y$ can be of order unity, but we always assume $x \ll 1$.
%
The next order in the multipole series is smaller by a factor 
$x^{2}$~\cite{VanDeHulst:1981}.

A discussion of the validity of the dipole approximation can be found in 
Sec.\ref{s:check-dipole-approx} below. We simply note here that our
aim is to compare the theory to apertureless Scanning Near-field Optical Microscopy 
(SNOM) measurements where scattering probes with a size of one micron or smaller 
were used, clearly smaller than the wavelength in the near infrared range 
(whence $x \ll 1$).
The partial screening of the field inside the particle (skin effect) is in fact
taken into account in the polarizabilities~(\ref{eq:Mie-alpha}, \ref{eq:Mie-beta})
via the parameter $y$, essentially the ratio between particle size and
skin depth in the material.
%

Note that our work here is limited to particles with an isotropic polarizability, 
but can be generalized
straightforwardly. For the more complicated case of spheroids, see 
Refs.\cite{Biehs:2010kp,Huth:2010fg}. 
Since many experimental devices use tungsten tips, we consider tungsten as material. 
We study the dressed polarisabilities for two spherical tip sizes (100 nm and 500 nm radii) and above three materials: SiC and SiO2, being both dielectrics, and gold.
The dielectric functions are taken from tabulated data~\cite{Palik:1985}.

\subsection{Contributions to the dressed polarizabilities}

Let us consider a tungsten sphere of 100 nm radius located 200 nm above the sample
surface. This configuration allows to observe multiple interactions with the surface without being in a distance regime where multipolar interactions are expected.

\begin{figure}
\begin{center}
\includegraphics[width=12cm]{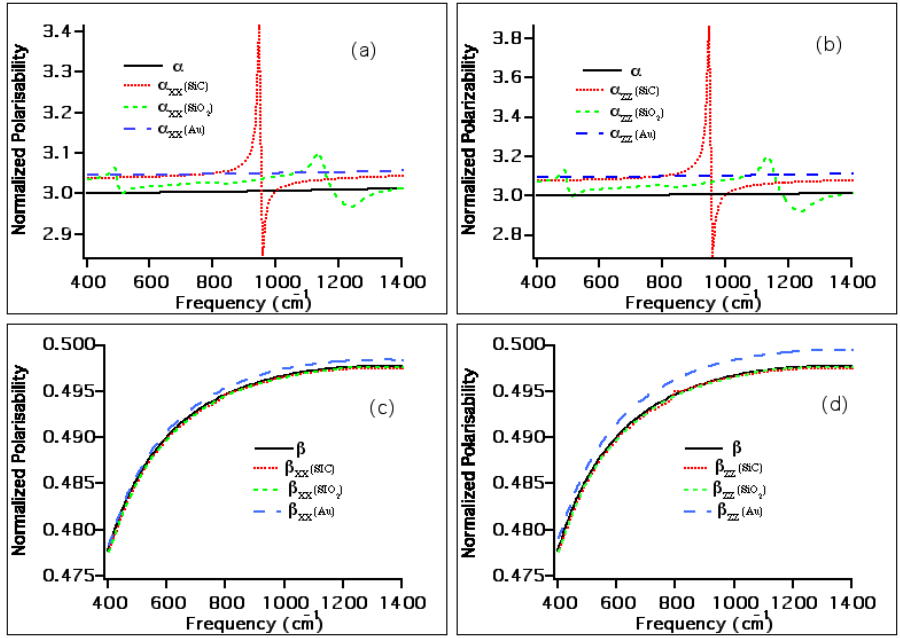}
\caption{Dressed polarizability tensor of a 100nm-radius tungsten sphere 
above a plane surface of SiC, SiO$_{2}$ or gold. The sphere center is at
a distance $z = 200\,{\rm nm}$ from the surface. 
Panels (a,b): electric polarizability $\alpht^{EE}$, parallel and 
perpendicular components.
Panels (c,d): magnetic polarizability $\bett^{HH}$.
All curves show absolute values, normalized to the volume of the particle.
The SiC surface shows a surface phonon polariton resonance at
$948\,{\rm cm}^{-1}$.
}
\label{aleff100200}
\end{center}
\end{figure}

In the case of a dipolar particle close to a plane interface, five different contributions to the polarizabilities are identified: the parallel electric ($\alpha^{EE}_{xx}$), perpendicular electric ($\alpha^{EE}_{zz}$), parallel magnetic ($\beta^{HH}_{xx}$), perpendicular magnetic ($\beta^{HH}_{zz}$) and crossed polarizability ($\alpha^{EH}_{xy}$ or $\beta^{HE}_{xy}$). The first four contributions are plotted in Fig.~\ref{aleff100200}. The crossed polarizability is not represented since it is two orders of magnitude smaller than the other contributions.
 
We observe that both perpendicular and parallel dressed electrical polarizabilities are quite different from the bare polarizability of the single particle. In particular, resonant features appear close to the frequencies of the SiC and SiO$_2$ surface phonon polariton mode. This is not surprising since the single-interface Green tensors appear in the dressed polarizabilities. The latter involve the reflection coefficients which diverge at the plasmon resonance. This is particularly pronounced at small distances from the interface. Thus the dressed polarizabilities are greatly affected in the near field and close to surface resonances. In contrast, no peak is seen in the spectra of gold dressed polarizabilities since gold does not exhibit resonances in the studied frequency range (mid-infrared). 
Very differently from the electric case, the magnetic polarizability is only weakly modified by the dressing. 
The main reason is that the bare magnetic polarizability shows a different 
scaling at small radius $R_t$. While the electrical polarizability behaves 
as $R_t^3$  (Clausius-Mossotti limit), the magnetic is proportional to 
$(R_{p}^{5} / \lambda^{2}) \left(\epsilon -1 \right)$ \cite{Chapuis:2008kcb}. 
In the present case, as the particle radius is much smaller than the wavelength, the magnetic polarizability is smaller; this implies that the dressing correction is also smaller, in particular for the dielectrics SiC and 
SiO$_2$. 
The correction is much more significant for gold because of its large dielectric 
constant in the infrared. This is a consequence of the fact that magnetic near fields 
are stronger at metals than above dielectrics.
 
\begin{figure}
\begin{center}
\includegraphics[width=12cm]{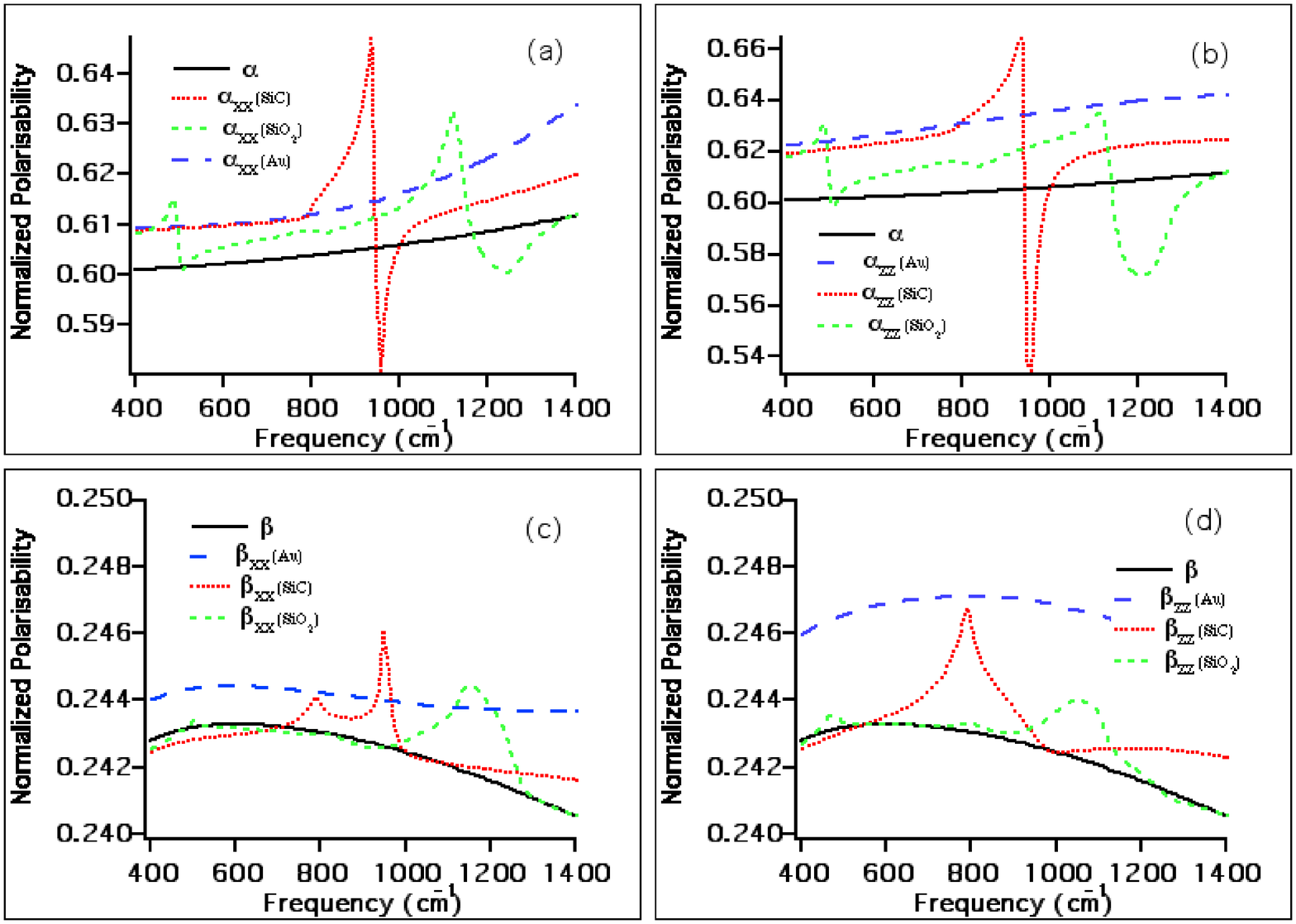}
\caption{
Dressed polarizabilities. Same as Fig.~\ref{aleff100200}, but for a larger
particle: radius $R_t = 500\,{\rm nm}$, distance $z = 1\,\mu{\rm m}$.
}
\label{aleff5001000}
\end{center}
\end{figure}

Let us now study the case of a larger particle (500 nm) at $1\,\mu{\rm m}$ above the sample. The dressed polarizabilities are represented in Fig. \ref{aleff5001000}. Here again crossed polarizabilities are not represented since they are smaller by two orders of magnitude.
We note significant corrections for both dressed electric and magnetic polarizabilities. These corrections are once again quite prominent around surface resonance frequencies.
The impact on the magnetic polarizabilities is now much more striking than for small particles due to the fact that here $R_t / \lambda$ is larger. We will see in the next section that the different behavior at $1\,\mu{\rm m}$ is due to the onset of
retardation.
This can be noticed by analyzing carefully the dressed polarizabilities.

\subsection{Asymptotic expressions}

We now simplify the dressed polarizabilities in order to get simpler  
expressions valid in the near field regime. We first start with Green 
tensors $\green^{EE}_R(\rv_t,\rv_t)$ and $\green^{HH}_R(\rv_t,\rv_t)$ that 
appear as integrals over the parallel wavevector 
[Eqs.(\ref{eq:ee-and-hh-Green}--\ref{eq:HHzz-Green})].

\begin{figure}
\begin{center}
\includegraphics[width=7cm]{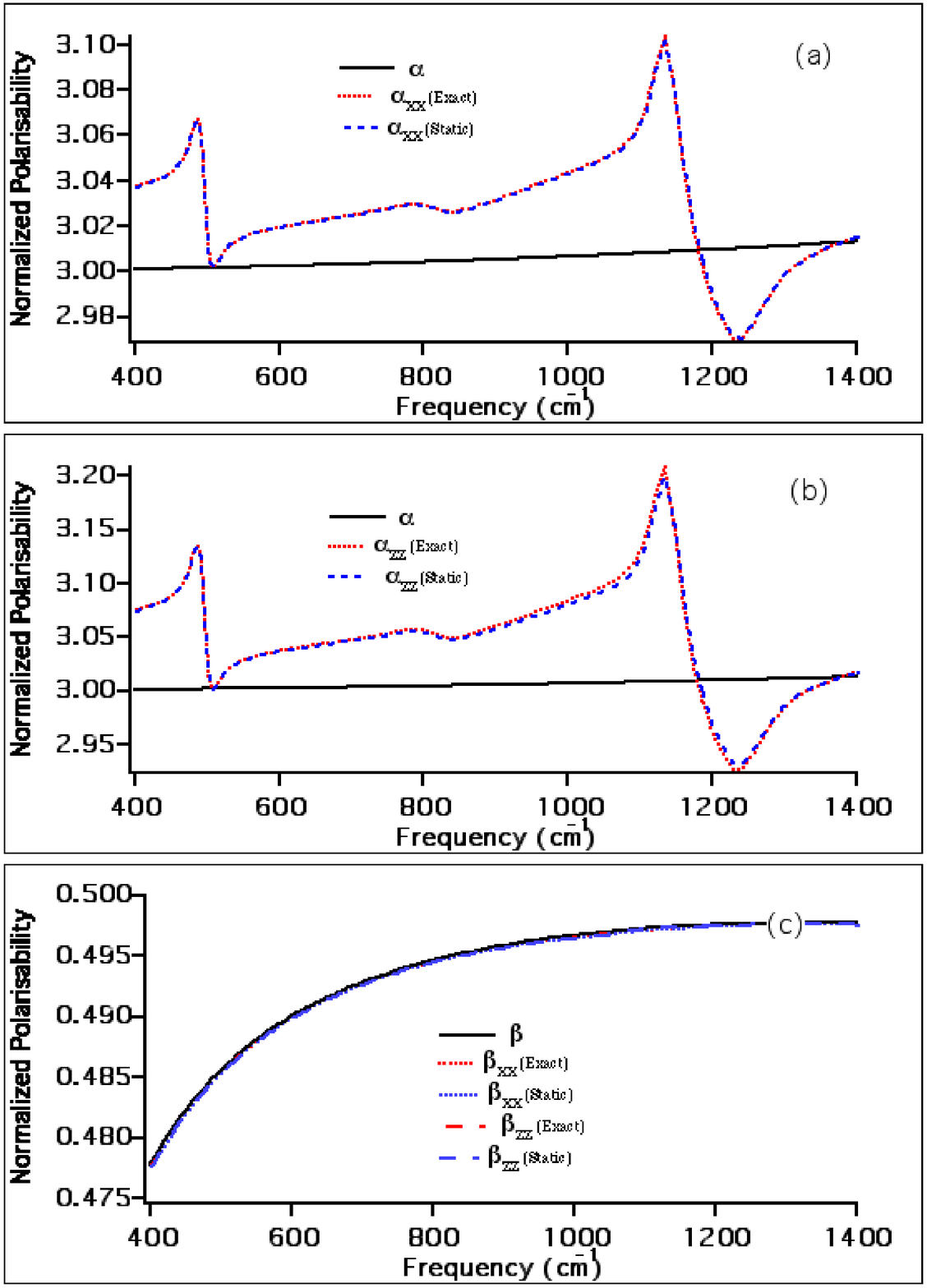}
\includegraphics[width=7cm]{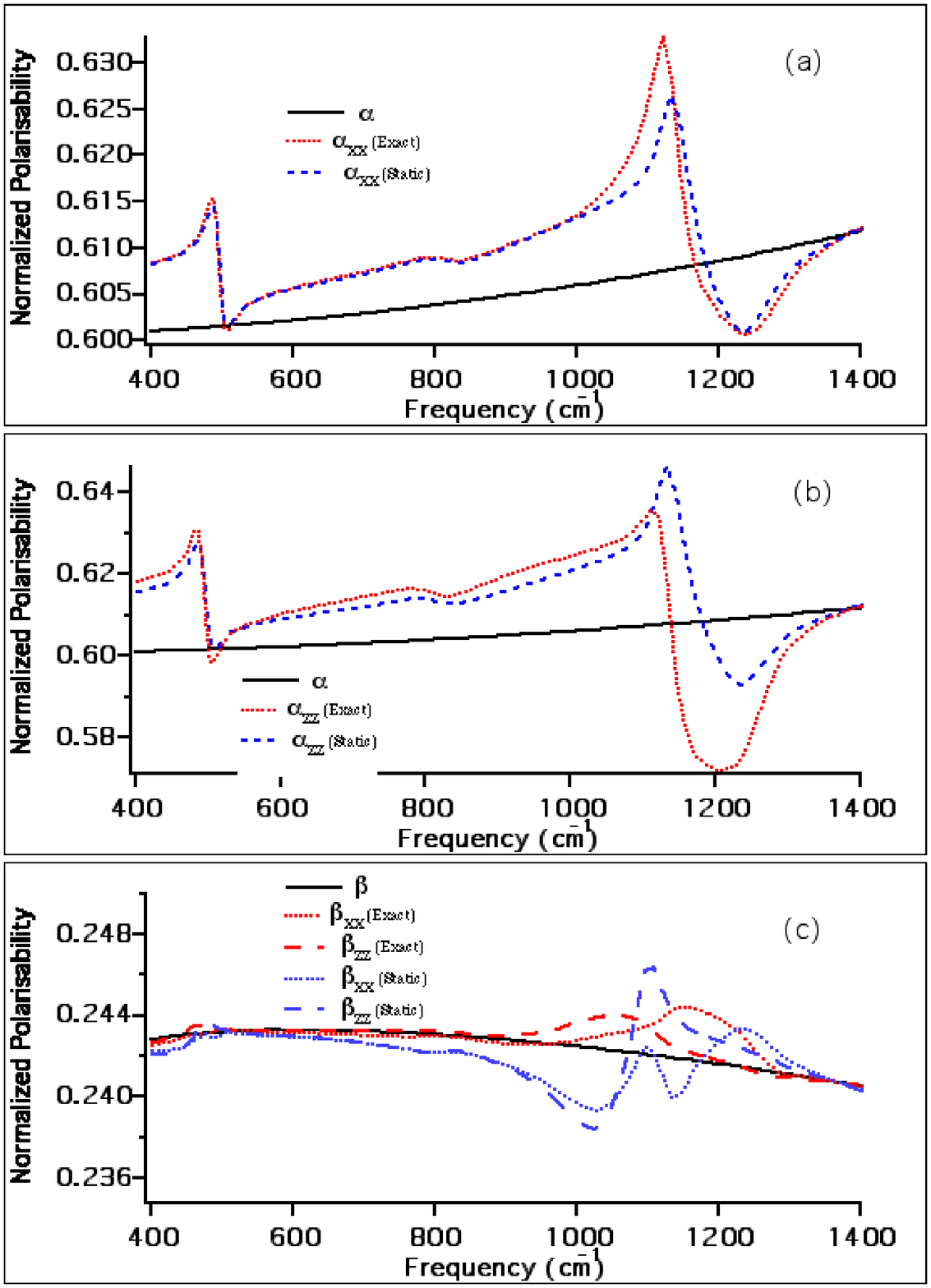}
\end{center}
\caption{Accuracy of the electrostatic approximation (extreme near field limit,
Eqs.(\ref{alphaeffapprox}, \ref{eq:beta-approx})) for the dressed polarizabilities. 
Left panel: small sphere with 100nm radius at $z=200\,{\rm nm}$.
Right panel: 500 nm radius at $z = 1\,\mu{\rm m}$.
The panels (a, b) give the parallel and perpendicular components of the 
electric polarizability, panel (c) the magnetic polarizability. 
In all cases, the values are normalized to the particle volume and plotted in
absolute value. The sphere is made from tungsten, the surface is SiO$_2$.
}
\label{compkeil}
\end{figure}

We suppose 
that the distance $z_t$ is much smaller than the wavelength, so that the 
exponential $e^{ 2 i \gamma z_t }$ allows for large values of $K$ to contribute
to the integrals~(\ref{eq:HExy-Green}--\ref{eq:HHzz-Green}).
This is valid in a regime sometimes called ``extreme near field"~\cite{Henkel:2000tr}.
In this regime, 
$r^p \approx (\epsilon - 1) / (\epsilon + 1)$ and 
$r^s \approx (\epsilon - 1) k_0^2 / 4 K^2$. 
Integration over the parallel wavevector $K$ is then easy since 
$\gamma^2 = k_0^2 - K^2 \approx -K^2$ so that 
$e^{ 2 i \gamma z_t } \approx e^{ - 2 K z_t }$.
For corrections to this regime that may even appear in the near field, see
Ref.\cite{Chapuis:2008kca}. 
Adopting the extreme near field regime, the Green tensors become
\begin{equation}
\label{ }
\green^{EE}_R(\rv_t,\rv_t)\approx\frac{1}{32\pi \epsilon_0 \, z_t^3}
\frac{\epsilon-1}{\epsilon+1}
\left(\begin{array}{ccc} 1 & 0 & 0 \\0 & 1 & 0 \\0 & 0 & 2 \end{array}\right)
\end{equation}
and
\begin{equation}
\label{ }
\green^{HH}_R(\rv_t,\rv_t)=\frac{k_0^2}{8\pi z_t}\left(\begin{array}{ccc}\frac{1}{2}\left(\frac{\epsilon-1}{4} + \frac{\epsilon-1}{\epsilon+1}\right)& 0 & 0 \\0 & \frac{1}{2}\left(\frac{\epsilon-1}{4} + \frac{\epsilon-1}{\epsilon+1}\right) & 0 \\0 & 0 &\frac{\epsilon-1}{4}\end{array}\right)
\end{equation}
These expressions are the asymptotic approximations of the Green tensors. 
For the electric term, it is also known as the electrostatic 
limit, which means that retardation is not taken into account. 
The dressed polarizabilities hence take the form:
\begin{eqnarray}
	\label{alphaeffapprox}
\alpht^{EE} &\approx& \left(\begin{array}{ccc}\frac{\alpha}{1-\frac{\alpha(\epsilon-1)}{32\epsilon_0\pi z_t^3(\epsilon+1)}} & 0 & 0 \\0 & \frac{\alpha}{1-\frac{\alpha(\epsilon-1)}{32\epsilon_0\pi z_t^3(\epsilon+1)}} & 0 \\0 & 0 &\frac{\alpha}{1-\frac{\alpha(\epsilon-1)}{16\epsilon_0\pi z_t^3(\epsilon+1)}}\end{array}\right)
\\
	\label{eq:beta-approx}
\bett^{HH} &\approx& \left(\begin{array}{ccc}\frac{\beta}{1-\frac{\beta k_0^2}{16\pi z_t}\left(\frac{\epsilon-1}{4} + \frac{\epsilon-1}{\epsilon+1}\right)} & 0 & 0 \\0 & \frac{\beta}{1-\frac{\beta k_0^2}{16\pi z_t}\left(\frac{\epsilon-1}{4} + \frac{\epsilon-1}{\epsilon+1}\right)} & 0 \\0 &0 &\frac{\beta}{1 - \frac{\beta k_0^2}{32\pi z_t}(\epsilon-1)}\end{array}\right)
\end{eqnarray}
This approximation for the electric case is identical to previous work of Knoll and Keilmann\cite{Knoll:2000wm}. In Fig.~\ref{compkeil}, the example of a tungsten spherical particle above SiO$_2$ shows that the electrostatic approximation is very good for small particles at short distances. For a larger particle ($1\,\mu{\rm m}$ radius), deviations are visible between the full calculation and the extreme near field (electrostatic) approximation. This is due to retardation that is clearly not negligible even though we are at subwavelength distances here.

From Eq.(\ref{eq:beta-approx}),
the dressing correction to the magnetic polarizability vanishes if retardation 
is discarded due to the factor $k_0^2$ in the denominators. This is illustrated
in Fig.~\ref{compkeil} for a small particle $R_t = 100\,{\rm nm}$. The extreme near field
approximation breaks down as we take a radius $R_t = 500\,{\rm nm}$ because of the
scaling $\sim R_t^5$ of the polarizability.
The calculations of the following sections are therefore based on the full Green
tensors, as given by the integrals in Eqs.(\ref{eq:HHxx-Green},\ref{eq:HHzz-Green}).

\section{Radiation scattered into the far field}

\begin{figure}
\begin{center}
\includegraphics[width=6.5cm]{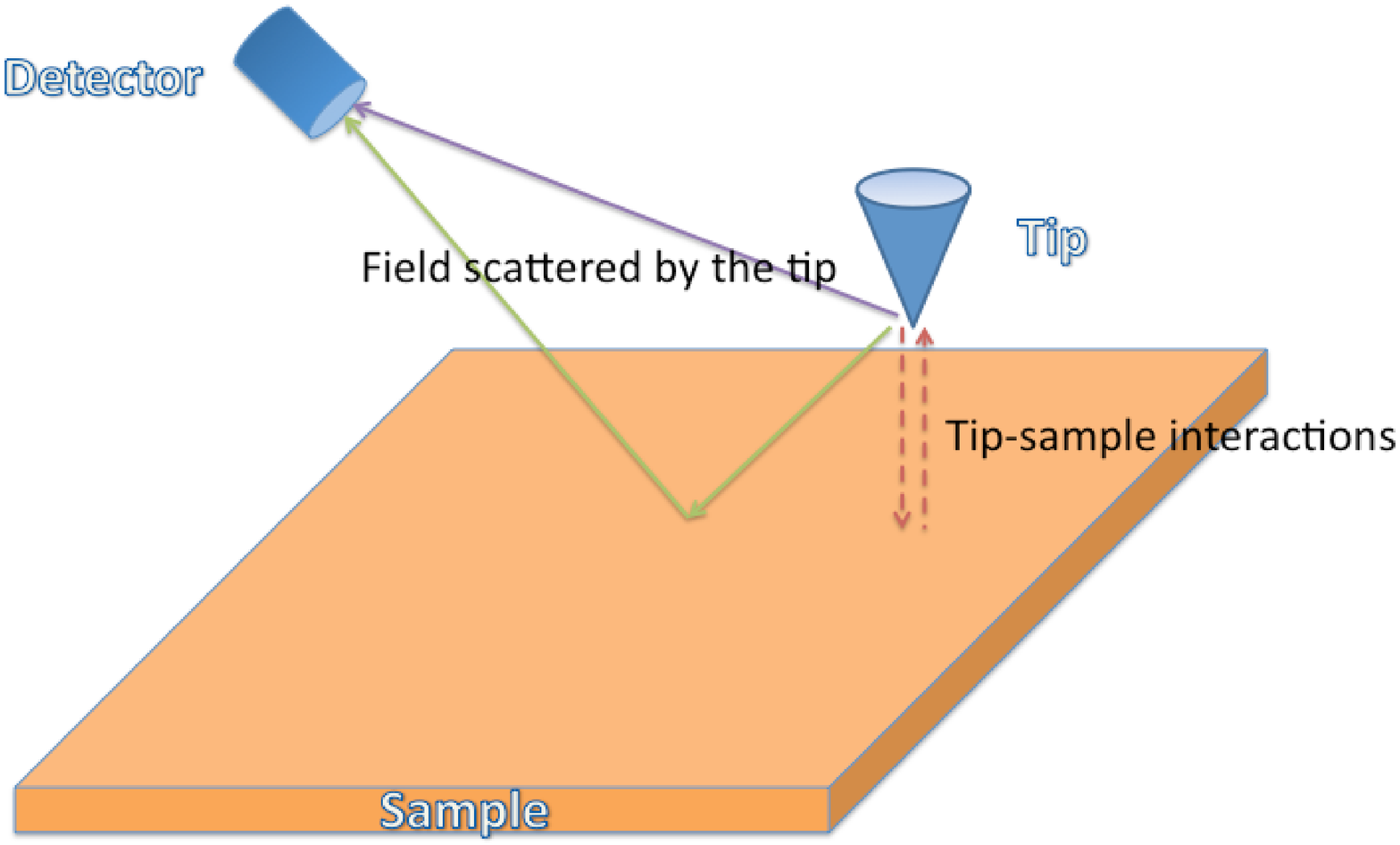}
\includegraphics[width=6.5cm]{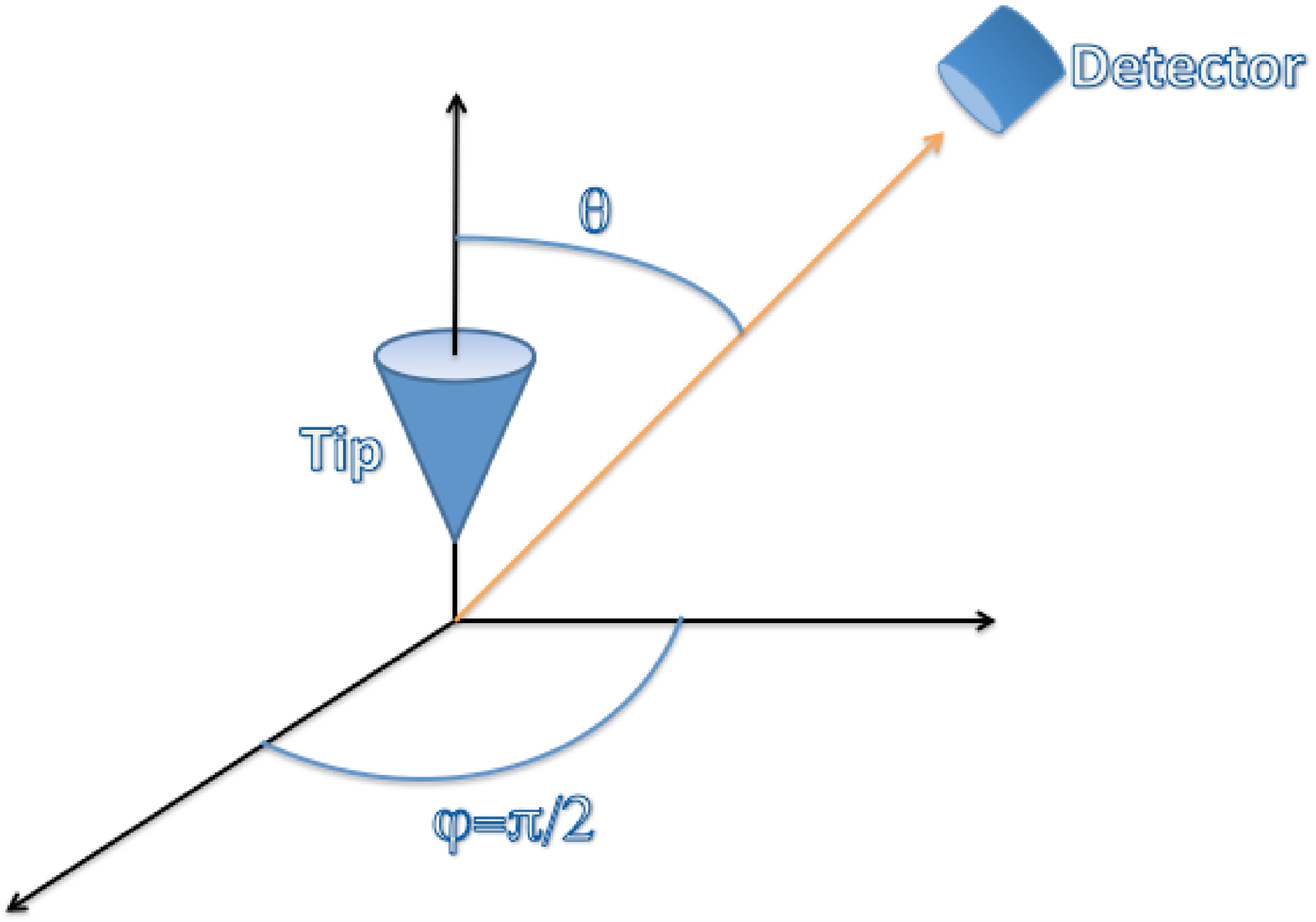}
\caption{Detection system scheme (left). Tip and detector position (right).}
\label{detscheme}
\end{center}
\end{figure}

The spherical particles studied so far provide a simple model for a SNOM tip. 
This kind of setup aims at detecting the near field of a sample, 
at a distance much smaller than the wavelength. The tip scatters this 
field and converts it into radiation that propagates to a detector placed
in the far field. See Fig.~\ref{detscheme} for a sketch of the geometry
where the emission is collected around a direction parametrized by the 
spherical coordinates $\theta, \varphi$.

\subsection{Signal at the detector}

The local electromagnetic field ${\bf E}^0( \rv_t ), {\bf H}^0( \rv_t )$
at the tip induces in the latter electric and magnetic dipole moments,
as described by the polarizabilites discussed before. The far field 
at the detector (distance $R$ from the tip) is just the electromagnetic 
radiation of these dipoles,
taking into account the reflection at the interface. The signal is
calculated in the far-field (Fraunhofer) approximation when 
$R$ is large enough compared to the size $R_d$ of the scatterer:
$R \gg R_d^2 / \lambda$. In this limit, the field at the detector can be 
considered a plane wave so that its power (averaged over one period) 
becomes
%
\begin{equation}
\label{signal1}
\left< S^d(\om) \right>
=\frac{\epsilon_0c}{2}|\E^d( \omega )|^2 r^2 d\Omega
\end{equation}
where $d\Omega$ is the solid angle subtended by the detector.
Let us introduce the unit vector $\uv_d$ pointing from the tip to the
detector and decompose it as
$\uv_d = \hat\uv_{d\parallel} \sin\theta + \hat{\bf z} \cos\theta$ where 
$\hat{\bf z}$ is the outward unit normal to the surface and the angle
$\theta$ is shown in Fig.~\ref{detscheme}. We define 
$\uv_{d}^{-} = \hat\uv_{d\parallel} \sin\theta  - \hat{\bf z} \cos\theta$ 
(the direction of a ray from tip to surface before reflection)
and the
(non-normalized) polarization vectors
%
$\shat_d = \hat \uv_{d\parallel}\times\ev_z$, 
$\ppp_d = \uv_d\times\shat_d$, 
and $\ppm_d = -\uv_d^-\times\shat_d$. 
%
These vectors and the Fresnel coefficients $r^{s,p}( \theta )$, evaluated for waves
at the angle $\theta$, are the building blocks for two tensors 
$\gamperp^{E,H}( \uv_d )$ that give the classical expression for the
detector field ${\bf E}^d$:
%
\begin{eqnarray}
\gamperp^E(\uv_d) &=& \shat_d\shat_d+\ppp_d\ppp_d
	+ (\shat_d r^s( \theta ) \shat_d+\ppp_d r^p( \theta ) \ppm_d)e^{i \phi} 
\\
\gamperp^H(\uv_d) &=& -\shat_d\ppp_d+\ppp_d\shat_d
	+ (-\shat_d r^s( \theta ) \ppm_d+\ppp_d r^p( \theta ) \shat_d) e^{i \phi}
\\
\E^d & = & \frac{\mu_0\omega^2}{4\pi}\frac{e^{ikR}}{R}
\left[ \gamperp^E( \uv_d ) 
\alpht^{EE}\E^0(\rv_t)\nonumber 
+
\frac{ 1 }{ c } \gamperp^H( \uv_d )
\bett^{HH}\Hv^0(\rv_t)
\right]
\end{eqnarray}
where $\phi = 2 k_0 z_t \cos\theta$ is the phase difference between the
`direct ray' from tip to detector and the ray once reflected at the surface.
%
%
We thus find the following expression for the detector signal~(\ref{signal1}):
\begin{equation}
	\label{eq:signal-and-Gamma}
\left< S^d( \omega ) \right> = 
\frac{\mu_0\om^4d\Omega}{32\pi^2c}
\sum_{i,j,k} \left(
\Gamma^E_{ij} \alpha^{EE}_{jj} E^0_j(\rv_t) 
+ \frac{ 1 }{ c }
\Gamma^H_{ij} \beta^{HH}_{jj} H^0_j(\rv_t)
\right)
\left(
\Gamma^{E*}_{ik} \alpha^{EE*}_{kk} E^{0*}_k(\rv_t) 
+ \frac{ 1 }{ c }
\Gamma^{H*}_{ik} \beta^{HH*}_{kk} H^{0*}_k(\rv_t)
\right)
\end{equation}
where we have used that the polarizabilities are diagonal, even when dressed.
This is a bilinear combination of electromagnetic field components, and
obviously not simply proportional
to the electromagnetic energy density at the tip.

\subsection{Expressions of the tensors $\gamperp^E$ and $\gamperp^H$}

For definiteness, we consider a situation where the detector 
is placed in the $yz$ plane, at an angle $\theta$ to the surface normal,
see Fig.~\ref{detscheme}.
%
The tensors in Eq.(\ref{eq:signal-and-Gamma}) then become
\begin{equation}
\label{ }
\gamperp^E=\left(\begin{array}{ccc}
1+r^s(\theta)e^{i \phi} & 0 & 0\\
0 & \cos^2\theta(1-r^p(\theta)e^{i \phi}) & -\sin\theta\cos\theta(1+r^p(\theta)e^{i \phi}) \\
0 & -\sin\theta\cos\theta(1-r^p(\theta)e^{i \phi}) & \sin^2\theta(1+r^p(\theta)e^{i \phi})
\end{array}\right)
\end{equation}
and
\begin{equation}
\label{ }
\gamperp^H=\left(\begin{array}{ccc}
0 & \cos\theta(1-r^s(\theta)e^{i \phi}) & -\sin\theta(1+r^s(\theta)e^{i \phi})\\
-\cos\theta(1+r^p(\theta)e^{i \phi}) & 0 & 0 \\
\sin\theta(1+r^p(\theta)e^{i \phi}) &0 & 0\end{array}\right)
\end{equation}
where the interference between the direct and reflected rays is manifest.
We therefore expect to see a signal that oscillates when the distance
$z_t$ is comparable to the wavelength.

\section{Application to apertureless SNOM experiments}

\subsection{Coherently excited surface polariton}

We now calculate the signal detected by an apertureless SNOM above a material supporting surface modes. Typical materials are metals \cite{Raether:1988ty} and 
more generally all materials with a dielectric constant smaller than $-1$. 
Surface polaritons are electromagnetic modes bound to a planar surface
and only exist for $p$ (or TM) polarization. 
They can be found by looking for a pole of the reflection amplitude $r^p$. 
In the coordinates of 
Fig.~\ref{detscheme}, the magnetic and electric fields of the surface mode 
propagating in the $y$-direction are given by
\begin{equation}
\label{ }
\Hv^0(\rv) = 
H_0 
\left(\begin{array}{c}1 \\0 \\0\end{array}\right)
e^{i(K y + \gamma z)}
,\qquad
\E^0(\rv) = 
\frac{ H_0 }{ \omega\epsilon_0 }
\left(\begin{array}{c}0 \\ -\gamma \\ K \end{array}\right)
e^{i(Ky + \gamma z)}
\end{equation}
where the pair $(K, \omega)$ satisfies the surface mode dispersion 
relation~\cite{Raether:1988ty}. This leads to $\Im \gamma > 0$ 
(evanescent mode) and a complex $K$ if we force $\omega$ to be real. 
The detector signal becomes
\begin{eqnarray}
\left<S^d( \omega )\right> & = & \frac{\mu_0 \omega^4 d\Omega }{ 32\pi^2 c^3 }
|H_0|^2e^{-2 \Im(K) y_t}e^{-2 \Im( \gamma ) z_t} 
	\label{sigdet} 
\\
&& {} \times \left[
\cos^2\theta|1-r^p(\theta)e^{i \phi}|^2 |\alpha^E_{yy}|^2
	\frac{| \gamma |^2}{ k_0^2 }
+ \sin^2\theta|1+r^p(\theta) e^{i \phi}|^2 |\alpha^E_{zz}|^2
	\frac{|K|^2}{ k_0^2 }\right.
\nonumber\\
&& {} + |1+r^p(\theta)e^{i \phi}|^2|\beta^H_{xx}|^2
\nonumber\\
&& {} + 2\sin\theta\cos\theta \,
\Re\left( (1-r^p(\theta)e^{i \phi})(1+r^{p*}(\theta)e^{-i \phi})
	\alpha^E_{yy}\alpha^{E*}_{zz} \frac{ \gamma K^* }{ k_0^2 }
	\right)
\nonumber\\
&& {} + 2\cos\theta \, \Re\left(
	(1-r^p(\theta)e^{i \phi})(1+r^{p*}(\theta)e^{-i \phi})\alpha^E_{yy}\beta^{H*}_{xx}
	\frac{ \gamma }{ k_0 }
	\right)
\nonumber\\
&& {} + \left.
2\sin\theta \, \Re\left( 
	|1+r^p(\theta)e^{i \phi}|^2\alpha^E_{zz}\beta^{H*}_{xx}\frac{ K }{ k_0 }
	\right)
	\right]
\nonumber
\end{eqnarray}
where $\alpht^{E} = \alpht^{EE} / \epsilon_0$,
$\bett^{H} = \bett^{HH}$, both with dimension volume.

\begin{figure}
\begin{center}
\includegraphics[width=10cm]{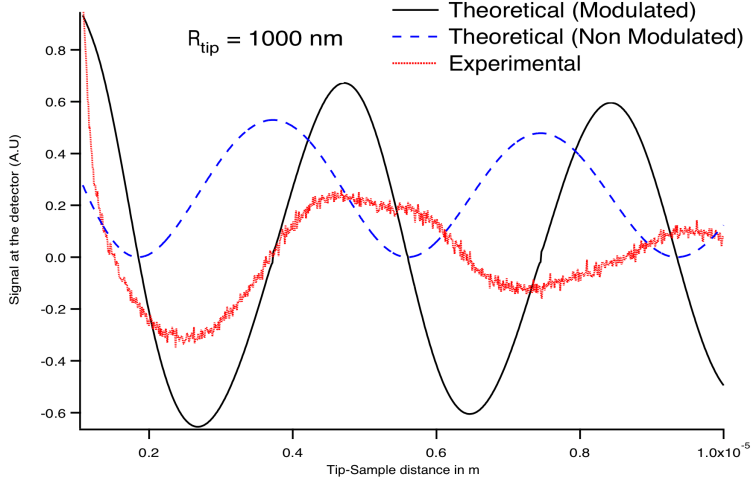}
\caption{TRSTM signal (theory and experiment) for a gold sample with
a surface plasmon mode excited at $\lambda = 7.5\,\mu{\rm m}$.
The excitation is performed with a quantum cascade laser cavity
integrated in the sample surface, as in the setup of Refs.\cite{Tetienne:2010gy, Babuty:dYmzb9en}.
%
We plot the intensity scattered into the far field at an angle
$\theta = 30^\circ$, as a function of tip-sample distance $z_t$.
Experiment (data points): 
detection through a Cassegrain objective (numerical aperture 0.5).
The tungsten tip (conical shape with half opening angle $\approx 20^\circ$) 
oscillates with an amplitude of 150\,nm,  
and the amplitude of the signal oscillating in phase with the tip is plotted.
Lines: theory, the tip is modeled by a tungsten sphere of 
radius $1\,\mu{\rm m}$. Solid line: simulation of the lock-in detection with the
same modulation amplitude. Dashed line: distance $z_t$ fixed. 
%
}
\label{Theoplasm}
\label{Expplasm}
\end{center}
\end{figure}

The detector signal from Eq.(\ref{sigdet}) is plotted in Fig.~\ref{Theoplasm}
as a function of distance $z_t$ (dashed curve).
A $1\,\mu{\rm m}$-radius tungsten sphere is taken as a model for a SNOM tip.
The oscillations in the lower panel illustrate the interference between the
direct and reflected rays. The period
is indeed around half the excitation wavelength $\lambda / 2 = 3.75\,\mu{\rm m}$.
In such experiments, the signal is rather weak and is extracted from the noise
with a lock-in amplifier, see Ref.~\cite{DeWilde:2006kt}. 
One modulates the tip-surface distance
and detects the signal at the modulation frequency or its second harmonic
(typical frequencies are in the kHz range).
The signal calculated for this modulation technique is also shown in 
Fig.~\ref{Theoplasm} (black solid line). As expected, the lock-in signal 
qualitatively arises from the derivative of the signal at fixed distance.
The modulation amplitude is rather large
($150\,{\rm nm}$) and at the shortest distances shown in the plot, the dipole
approximation is probably no longer reliable.


We have also plotted in Fig.~\ref{Expplasm} the experimental TRSTM data obtained
as described above. The experimental setup is similar to the one used in
Refs.\cite{Tetienne:2010gy, Babuty:dYmzb9en}. 
Here the plasmon is excited in an integrated plasmonic device by end-fire coupling 
between one end
facet of a quantum cascade laser cavity at 7.5 $\mu$m and the surface of a planar 
gold strip. The wavelength is very much below
the plasma resonance of gold, therefore the surface plasmon mode 
penetrates significantly outside the surface.
From the Drude parameters for gold, we estimate a extension into the vacuum 
of $1 / \Im \gamma \sim 60\,\mu{\rm m}$.
The overall behaviour of the experimental signal
is quite close to the theoretical prediction. 
The experimental curve decays somewhat faster with distance. We attribute this
to the different signal collection: indeed, the detector involves a Cassegrain 
objective that collects a finite range of angles $\theta$. This 
averages over interference fringes with different periods (recall the 
$\cos\theta$ in the exponential in Eq.(\ref{sigdet})). The modelling of
the tip by a sphere may also reach its limits here.
By comparing the experimental signal to the theory calculated for a spherical
particle, we find an effective tip radius of around $1\,\mu{\rm m}$. This is
about the same value as recently found \cite{Babuty:dYmzb9en}. The volume of
this spherical tip matches roughly the effective scattering volume of the tip
whose estimation we discuss in Sec.\ref{s:check-dipole-approx}, given the tip's
conical shape.

\subsection{Thermal emission from a heated surface}

\subsubsection{Thermal fields above the sample}

When the electromagnetic field above the surface is only the one due to thermal emission from the sample surface, several simplifications occur. The TRSTM signal of
Eq.(\ref{eq:signal-and-Gamma}) is calculated using correlation functions in
thermal equilibrium for the non-perturbed fields ${\bf E}^0( \rv_t )$, 
${\bf H}^0( \rv_t )$ \cite{Joulain:2005ih,Volokitin:2007el,Joulain:2010bq}.
Most of the cross correlations functions 
%
vanish due to the fact that thermal currents decorrelate for different directions \cite{Joulain:2005ih}. Due to rotational symmetry around the $z$-axis, we are
left with
$\langle|E^0_x(\rv_t)|^2\rangle = \langle|E^0_y(\rv_t)|^2\rangle$, 
$\langle|E^0_z(\rv_t)|^2\rangle$,
$\langle|H^0_x(\rv_t)|^2\rangle = \langle|H^0_y(\rv_t)|^2\rangle$,
$\langle|H^0_z(\rv_t)|^2\rangle$,
and
$\langle E^0_x( \rv_t ) H^{0*}_y( \rv_t ) \rangle = 
- \langle E^0_y( \rv_t ) H^{0*}_x( \rv_t ) \rangle$.
 
In these conditions, the signal at the detector in a direction making an angle $\theta$ with the $z$ axis reads
\begin{eqnarray}
\left< S^d( \omega ) \right> & = & \frac{ \omega^4 }{32\pi c^3 }d\Omega
\left\{
\left( \cos^2\theta|1-r^p(\theta)e^{i \phi}|^2
	+ |1+r^s(\theta)e^{i \phi}|^2
\right)
	|\alpha^E_{xx}|^2
	\langle \epsilon_0 |E^0_x(\rv_t)|^2 \rangle 
\right.\nonumber\\
&& {} + \sin^2\theta|1+r^p( \theta ) e^{i \phi}|^2 |\alpha^E_{zz}|^2
\langle \epsilon_0 |E^0_z(\rv_t)|^2 \rangle
\nonumber \\
&& {} + \left(
	\cos^2\theta|1-r^s(\theta)e^{i \phi}|^2
	+ |1+r^p(\theta)e^{i \phi}|^2
	\right)
	|\beta^H_{xx}|^2
\langle \mu_0 |H^0_x(\rv_t)|^2 \rangle
\nonumber\\
&& {} + \sin^2\theta|1+ r^s( \theta ) e^{i \phi}|^2 
	|\beta^H_{zz}|^2
\langle \mu_0 |H^0_z(\rv_t)|^2 \rangle
\\
&& {} + 2\cos\theta\,\Re\left[
	\alpha^E_{xx}\beta^{H*}_{xx}
	\langle E^0_x(\rv_t) H^{0*}_y(\rv_t) / c \rangle
	\right.\nonumber\\
&&{} \times \left.\left.\left(
	(1+ r^s( \theta ) e^{i \phi})(1- r^{s*}( \theta ) e^{-i \phi})
	+ (1- r^p( \theta ) e^{i \phi})(1+ r^{p*}( \theta ) e^{-i \phi})
	\right)\right]\right\} \nonumber
\end{eqnarray}
The thermal radiation above a plane interface is well known in the 
literature and yields the field correlations 
\cite{Joulain:2005ih,Volokitin:2007el,Dorofeyev:2011bg}:
\begin{eqnarray}
\left<|E^0_x(\rv_t)|^2\right> & = & 
\frac{\mu_0\omega \Theta(\omega,T) }{2\pi^2}
\Im\left(
i\int_0^\infty\frac{K dK}{ \gamma }
\left[1 + r^s e^{2i \gamma z_t} + \frac{ \gamma^2 }{ k_0^2 } (1 - r^p e^{2i \gamma z_t})
\right]\right) 
\label{ex2}
\\
\left<|E^0_z(\rv_t)|^2\right> & = & 
\frac{\mu_0\omega \Theta(\omega,T) }{\pi^2}
\Im\left(
i\int_0^\infty\frac{K^3 dK}{ \gamma k_0^2 }
(1 + r^p e^{2i \gamma z_t})
\right)
\\
\left<|H^0_x(\rv_t)|^2\right> & = &
\frac{\epsilon_0\omega \Theta(\omega,T) }{2\pi^2}
\Im\left(
i\int_0^\infty\frac{K dK}{ \gamma }
\left[1 + r^p e^{2i \gamma z} + \frac{ \gamma^2 }{ k_0^2 } (1 - r^s e^{2i \gamma z})
\right]\right)
\\
\left<|H^0_z(\rv_t)|^2\right> & = & 
\frac{\epsilon_0\omega\Theta(\omega,T)}{\pi^2}
\Im\left(
i\int_0^\infty\frac{K^3 dK}{ \gamma k_0^2 }
(1 + r^s e^{2i \gamma z})
\right)
\\
\left< E^0_x(\rv_t) H^{0*}_y(\rv_t) \right> &=& 
\frac{ \Theta(\omega,T) }{ 4\pi^2 }
\left[ \int_0^\infty \frac{2K \gamma dK}{ |\gamma| }
\Re\left(	
	\frac{ \gamma }{ |\gamma| }(1 + r^s e^{2i \gamma z_t}
		- r^p e^{2i \gamma z_t})
	\right)\right]
\label{exhy}
\end{eqnarray}
We denote $\Theta(\om,T) = \hbar \omega/[\exp[\hbar\omega/(k_BT)]-1]$ 
the mean thermal energy of an oscillator with angular 
frequency $\omega$, $\hbar$ and $k_B$ are the Planck and Boltzmann constants,
and $T$ is the temperature. The integrals are somewhat similar to
those in Eqs.(\ref{eq:HExy-Green}--\ref{eq:HHzz-Green}), being plane-wave
expansions; we use the same notation as there.

If the substrate and the particle dipole materials are known, the signal at the detector can be calculated from these formulas. 
If the signal is divided by the mean energy of an oscillator 
$\Theta(\om,T)$, it only depends on the particle polarizability, 
the tip-sample distance and the surface optical properties. The polarizability can
be extracted from reference experiments at another surface.
Note that the dressed polarizability and the thermal electromagnetic field are 
closely related. Indeed, both involve the system's Green tensor taken at the 
tip position. 
As a consequence, if the thermal near field shows a resonance at a given frequency, 
it is likely that there will be one in the dressed polarizability at the same 
frequency. This entails a difficulty for the interpretation of the signal at 
the detector. Indeed, the electromagnetic field correlations are closely related 
to the electromagnetic local density of states (EM LDOS) \cite{Joulain:2005ih}, at least to a projected LDOS. Thus, $\langle |E_x|^2 \rangle$ is related to the parallel electric LDOS whereas $\langle |H_z|^2 \rangle$ is related to the perpendicular magnetic LDOS. Relating the detector signal to the LDOS would be very interesting, since it would give a way to detect this quantity like electronic tunneling microscopy does it for the electronic LDOS \cite{Tersoff:1985wm}. Unfortunately, it is not possible to find a simple and universal relation between the two quantities. 
We shall see, however, that in specific situations, one contribution may be 
the leading term, so that the projected EM LDOS and the detected signal become 
proportional.

\subsubsection{Probing a polar material}

\begin{figure}
\begin{center}
\includegraphics[width=12cm]{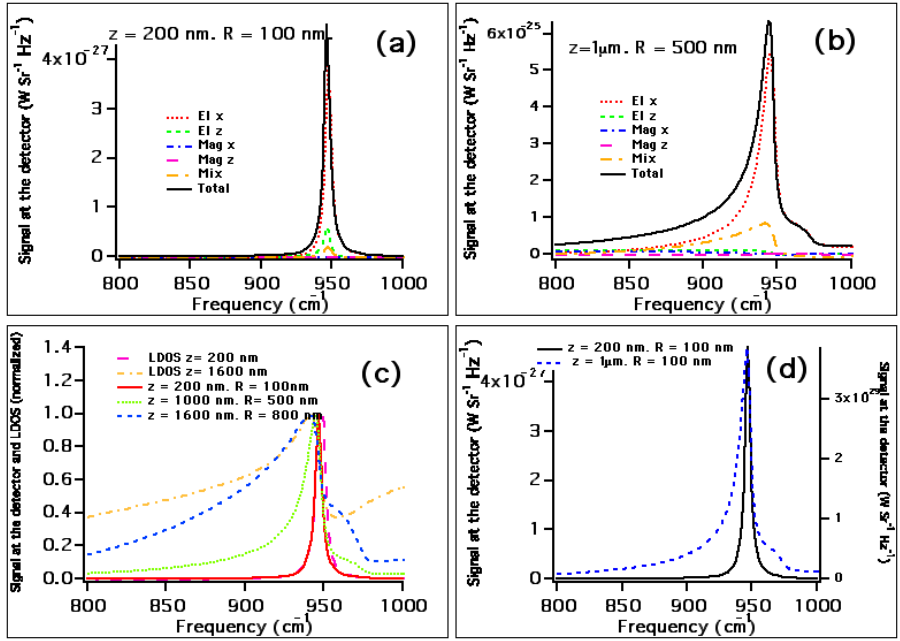}
\caption{Spectrum of thermal radiation scattered by a tip
(TRSTM signal). 
The tip is described by a tungsten sphere
above a SiC sample. The signal is detected in a direction making an angle of 30$^\circ$ with the vertical direction. 
(a) Signal and its contributions 
at height $z = 200\,{\rm nm}$. We take $R_t = z /2$ in panel (a--c).
(b) Signal 
at height $z = 1\,\mu{\rm m}$. 
(c) Comparison of the signal with the LDOS 
for $z 
= 200\,{\rm nm}$, $z 
= 1\,\mu{\rm m}$ 
and $z 
= 1.6\,\mu{\rm m}$.
(d) Signal spectrum for $R_{p} = 100\,{\rm nm}$ at two different distances 
$z = 200$ nm (left scale) and $z = 1\,\mu{\rm m}$ (right scale).
}
\label{sigdetWSiC}
\end{center}
\end{figure}

We now study the spectrum of the signal detected by a probe above SiC. As we have seen in the previous section, this signal strongly depends on the dressed polarizability and therefore on the tip size and the tip distance. To illustrate this, we show, in Fig.~\ref{sigdetWSiC}, the spectral signal detected by a tungsten probes above SiC with radii 100 nm and $1\,\mu{\rm m}$. It is observed that the emission peak is rather narrow for the small tip. This peak appears around the phonon polariton resonance frequency that is 948 cm$^{-1}$ for SiC.
We also note that for a small tungsten tip, the signal above SiC is dominated at short distance by electrical terms. 
%
Parallel and perpendicular field components contribute both significantly. 
These contributions show a slight spectral shift with respect to each other: this can be attributed to the different denominators in the dressed polarizability tensor of Eq.(\ref{alphaeffapprox}).

For a 500 nm tip at $1\,\mu{\rm m}$ from the interface, we note that the main contribution comes from the parallel electric field. Magnetic fields are more important here because the magnetic polarizability is larger. These magnetic terms do not contribute a lot because near the polariton resonance in a polar material, the energy density is dominantly electric. However, the mixed term, which involves both magnetic and electric fields (and both polarizabilities), is the second contribution to the signal and is therefore not negligible. 
Fig.~\ref{sigdetWSiC}(c) shows various signal spectra for three tips at (center-surface) distances equal to the their diameter. They are compared with the LDOS at $200\,{\rm nm}$ and $1.6\,\mu{\rm m}$. At 200 nm distance, the tip gives a signal similar to the electromagnetic LDOS: a well-defined peak appears around the polariton frequency. Its width is very similar to the LDOS, but the position is slightly shifted. When the tip size increases, the peak 
tends to shift and to broaden. 
Recent SNOM experiments based on the measurement of the near-field thermal emission using a tungsten tip seem to confirm this shifting and broadening, suggesting that the approximation of the tip by a simple spherical dipole is valid to some extent \cite{Babuty:dYmzb9en}.
Note also that this broadening, although less pronounced than for a large tip, also occurs when a small tip is retracted from the surface as it can be seen in Fig.~\ref{sigdetWSiC}(d). 


As shown in Fig.~\ref{sigdetWSiC}, the signal calculated with a small tip is very similar to the LDOS around the polariton resonance. At short distance, the signal is indeed dominated by the parallel electric contribution. This means that the signal mainly depends on $\alpha^{EE}_{xx}$ and on $|E^0_x|^2$. In the case of a thermal signal, this last quantity is proportional 
to the projected parallel electromagnetic LDOS. Therefore, a SNOM experiment detecting the thermal near field will measure the product of one of the dressed polarizabilities by a partial contribution to the EM LDOS. If the dressed polarizability has a flat spectral 
response (i.e. it varies only weakly with frequency), 
one can say that the signal at the detector is proportional to the projected EM LDOS. This is not strictly the case in the present situation where the polarizability can be increased by 10--20\% around the resonance 
(Fig.~\ref{aleff100200}):
%
the resulting signal is the product of two peaks at approximately the same frequency. As the value of the EM LDOS in the peak spectral band is about 100 to 1000 times its value outside this band, its multiplication by the dressed polarizability will give a peak that is not exactly the projected EM LDOS but which is representative of the LDOS.    

Consider now the results for the larger tip (radius $1\,\mu{\rm m}$). The signal scattered to the detector by such tips has a broader spectrum and the frequency corresponding to the emission spectrum is shifted to lower frequencies [Fig.~\ref{sigdetWSiC}(b)]. 
When the tip size increases, polarizability is shifted to lower frequency and 
broadened [compare Figs.\ref{aleff100200} and \ref{aleff5001000}]. At larger 
distances, the EM LDOS is also broadened [Fig.~\ref{sigdetWSiC}(c)]. 
The resulting signal is now the product of two peaks which are rather similar in shape but with a slight frequency offset. This product shows a maximum mid-way between the peaks of the polarizability and the LDOS. Under these conditions, the signal at the detector is a peak related to the EM LDOS, but is not strictly speaking giving the EM LDOS. However, one can see from Fig.~\ref{sigdetWSiC}(c) that the detected signal and the LDOS have a very similar shape.

\subsubsection{Probing a metallic sample}

\begin{figure}
\begin{center}
\includegraphics[width=12cm]{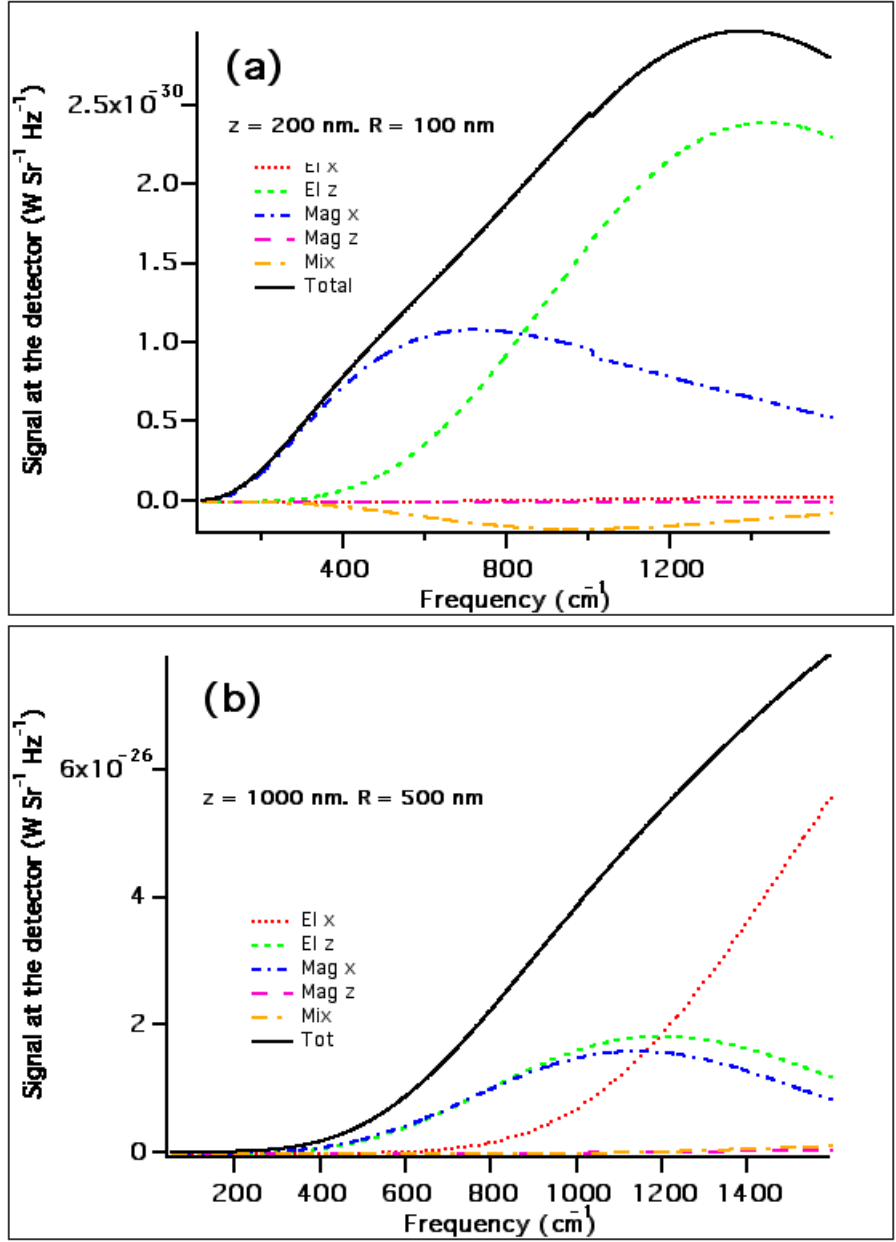}
\caption{Spectrum of TRSTM signal, as in Fig.~\ref{sigdetWSiC}, but above a 
gold sample. The tip is described by a tungsten sphere. The signal is detected in a direction making an angle of 30$^\circ$ with the vertical (a) Total signal and different contributions for a 100 nm radius sphere at $z=200\,{\rm nm}$ above surface. (b) Total signal and different contributions for a 500 nm radius sphere at $z=1\,\mu{\rm m}$ above the surface.}
\label{sigdetWAu}
\end{center}
\end{figure}

We now consider the example of a tip above a metal surface. Fig.~\ref{sigdetWAu} shows the signal scattered by a tungsten particle above a gold surface heated at 300 K. 
Note the significantly different spectral dependence and the reduced
magnitude, compared to the polar dielectric of the previous section.
For a small particle, the contribution to the signal is dominated by the magnetic term below 1000 cm$^{-1}$. Indeed, the magnetic energy is larger than the electrical energy in this spectral range where Au is highly reflecting. As seen in Fig.~\ref{aleff100200}, the magnetic polarizability has a rather flat response for small tips above 400 cm$^{-1}$. As a consequence, the signal at the detector is proportional to the parallel 
magnetic LDOS between 400 and 1000 cm$^{-1}$. 
Since the total EM LDOS is dominated by its magnetic contribution,
the signal detected in this spectral range is close to the EM LDOS. On the contrary, above 1000 cm$^{-1}$, the signal is dominated by the perpendicular electric contribution but the parallel magnetic and the mixed term are of the same order of magnitude even if smaller. For such situation, it is not correct to state that the signal is proportional to the EM LDOS although it is related to it. 
For a large tip, there is no spectral range where one component clearly dominates the contribution to the signal (except at high frequencies, but there the validity of the dipole approximation becomes questionable).
The detected signal cannot be considered as simply proportional to the EM LDOS.

\subsection{Modeling the tip by a dipolar particle}
\label{s:check-dipole-approx}

Considering the calculations made in this paper and the sharp tips in real experiments,
one may wonder why a conical tip is well described by a spherical particle. As a 
matter of fact, TRSTM experiments are in excellent agreement with this theoretical
model~\cite{Babuty:dYmzb9en}. This can be qualitatively understood with the
following argument. To be specific, we consider the detection of the thermal near field
above a sample and a sharp conical tip whose apex is touching the surface.
The field intensity (autocorrelation function) typically decays
like $1/z^3$ with the distance $z$ from the surface. A slice of the tip at height $z$ 
has a volume proportional to $z^2$ that contributes
to the scattered field, provided the field fully penetrates into the material. 
Therefore, the signal arising from this slice is proportional to 
$1/z^3\times(z^2)^2 \propto z$, favoring the tip's shaft rather than the apex.
Above a certain altitude, the field does not penetrate into the metallic tip, and the 
scattering volume of a slice is determined by its circumference, proportional to $z$.
This leads to a scattered intensity $\propto 1/z$, indicating that most of the signal 
arises from an optimal height proportional to the skin depth and depending on the
cone's opening angle. 
The skin depth for tungsten in the infrared around $10\,\mu{\rm m}$ is 
$\lambda/\sqrt{\epsilon}\sim 200\,{\rm nm}$. This means that even for tips touching 
the sample, the field scattered will arise from a distance of several hundreds of
nanometers. This argument makes it plausible why the simple model of a particle with 
relatively large size permits to
capture much of the phenomenon, as found by comparing to experimental data in
Ref.\cite{Babuty:dYmzb9en}. The same argument also holds for the situation of
a coherently excited surface mode, as illustrated by Fig.~\ref{Theoplasm} above:
indeed, the local field intensity then also decays
with distance, even though the dependence on $z$ differs. 

\section{Thermal emission of a nanoparticle}

We now consider a small dipolar particle heated to a temperature $T_t$. This particle radiates an electromagnetic field in all space
which can be detected in the far field. 
This configuration corresponds to recent experimental situations \cite{Jones:2012fx} where a heated tip is moved into the near field of a sample. Note that in this situation, 
the detector is also picking up the thermal emission coming from the sample,
as calculated in the preceding section. We focus in the following on the radiation 
from the tip; it could be isolated by suitable imaging and modulation techniques.
%
The tip signal has again contributions coming directly from the particle in straight line and undergoing one reflection at the interface. The signal at the detector reads:
\begin{eqnarray}
\label{ }
\left<S^d(\omega)\right>&=&\frac{d\Omega}{32\pi^3}\Theta(\om,T_t)\frac{\om^3}{c^3}\left[\Im[\alpha^E_{xx}]\left(|1+r^s(\theta)e^{i \phi}|^2+\cos^2\theta|1-r^p(\theta)e^{i \phi}|^2\right)\right.\nonumber\\
&+&\left.\Im[\alpha^E_{zz}]|1+r^p(\theta)e^{i \phi}|^2\sin^2\theta\right]
\end{eqnarray}
where $T_t$ is the tip temperature.
Here, the spectral dependence comes mainly from the dressed polarizability. 
The far field reflection coefficients $r^{s,p}( \theta )$ typically show 
no resonant features since surface modes appear in the evanescent sector. 
One thus expects that the detected spectra will follow polarizability variations that we have described in the preceding sections.

\begin{figure}
\begin{center}
\includegraphics[width=12cm]{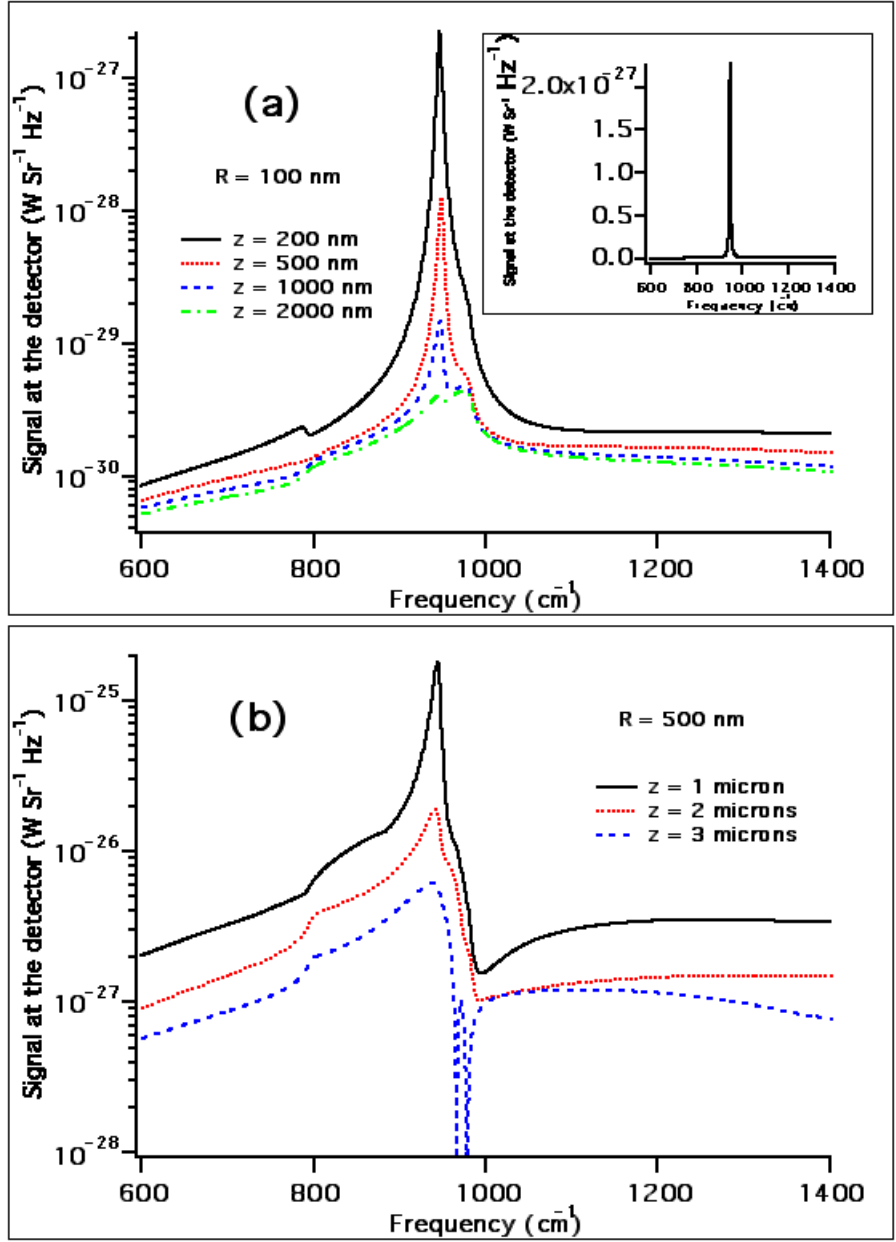}
\caption{Signal at the detector at an angle $\theta = 45 ^0$ to the surface 
normal (log scale). The tip is described by a tungsten sphere heated at 
$T = 300$ K and located at various distances to the interface. (a) Signal in log-scale for $R_t = 100\,{\rm nm}$,
for different tip-sample distances. Inset: signal for $R_t= 100\,{\rm nm}$ and 
$z= 200\,{\rm nm}$ in linear scale. 
%
(b) Signal for $R_t = 500 \,{\rm nm}$ and at distances $z = 1, 2, 3\,\mu{\rm m}$.}
\label{Chff}
\end{center}
\end{figure}

In Fig.~\ref{Chff}, we plot the signal detected in the far field in a direction making an angle of 45$^\circ$ with the vertical direction. The tip is modeled by a tungsten sphere (100 nm radius) heated at 300 K and held at various distances from a vacuum-SiC interface. 
Note that the tip sample distance is much smaller than the wavelength 
so that $\phi\ll1$.
%
When the particle-interface distance is 200 nm, the signal is peaked around the SiC surface resonance ($948\,{\rm cm}^{-1}$), similarly to the polarizability of a 100 nm-radius sphere. When the particle is retracted from the interface, the signal is reduced and the spectrum broadens. 
%
Above a certain distance, the dressing corrections to the polarizability become
negligible, and for tungsten, the spectral dependence becomes flat.
The spectral behaviour of the signal comes mainly from variations of the reflection coefficient. 
One notes that the signal is more important in the frequency range where SiC is known to be highly reflective i.e. between 
$850\,{\rm cm}^{-1}$ and $950\,{\rm cm}^{-1}$.
A similar behaviour is observed for a micron-sized tip except that the signal is broader at the minimum distance. This can easily be explained by inspecting the dressed polarizabilities of a 500 nm-radius sphere 
above a SiC interface where this broadening is also observed
(compare Figs.~\ref{aleff100200} and \ref{aleff5001000}). 

Note that for such situation, the detected signal can exhibit a peak very similar to the one observed in the energy density spectrum or in the EM LDOS. However, this peak is the signature of the dressed polarizability which exhibits a resonance at frequencies close to surface resonance. Even if such an experiment does not probe the EM LDOS, it can probe surface resonances if the tip is sufficiently close to the interface~\cite{Jones:2012fx}. 

\section{(Non)radiative cooling of a particle}


When a particle is heated, it exchanges energy with the environment and cools
down. In free space and assuming the 
surroundings at zero temperature, the spectral power lost by the particle
(temperature $T_t$, modelled by a dipole) is \cite{Joulain:2008tp}   
\begin{equation}
\label{ }
P(\om)=\frac{\om^3}{\pi^2 \epsilon_0 c^3}\Im[\alpha(\om)]
\Theta(\om,T_t)
\end{equation}
where $\Theta(\omega, T_t)$ is again the mean thermal energy of the dipole oscillator.
When the particle is close to a surface, the transferred power depends on the electromagnetic mode density at the particle position (EM LDOS) and can be much
larger than in free space because non-radiative channels (evanescent modes) open up.
%
This phenomenon is very similar to the Purcell effect when an atom or a molecule has its spontaneous emission rate modified inside a cavity or when approaching a surface. The cooling rate depends on the EM LDOS~\cite{BenAbdallah:2011tz} in a similar way as the spontaneous emission rate. 
The cooling rate formula given by Mulet et al.\cite{Mulet:2001kp} should be corrected at very close distances to take into account the dressed polarizability. We thus
find for the heat transferred from particle to sample
\begin{equation}
\label{eq:spectrum-cooling}
P(\om) = \frac{2}{\pi} \Theta(\om,T_t) 
\sum_{i=x,y,z} \Im(\alpha^E_{ii}) \Im[ G^{EE}_{ii}(\rv_t,\rv_t) ]
\end{equation}
This is easily generalized to particle and surroundings at temperatures $T_t, 
T_s > 0$, see Ref.\cite{Joulain:2008tp}.
We emphasize that the heat exchange is the observable that shows the ``cleanest'' 
connection to the EM LDOS which is itself given by the imaginary part of the
EM Green function.

\begin{figure}
\begin{center}
\includegraphics[width=12cm]{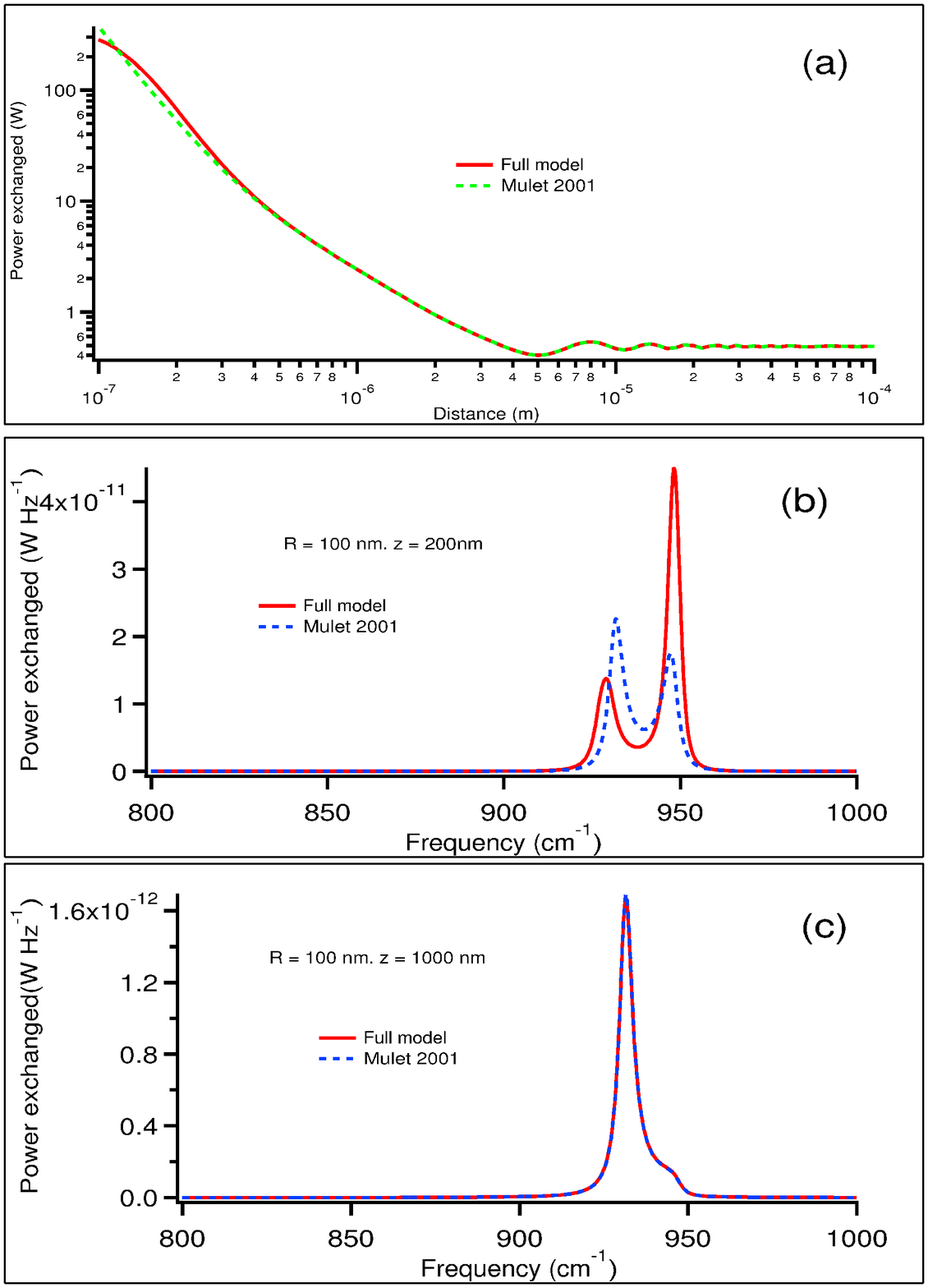}
\end{center}
\caption{(a) Total exchanged heat power between a 100nm radius SiC 
particle (300 K) above a SiC substrate ($T = 0$) versus distance. 
The two curves are based on the bare (``Mulet 2001'') and dressed 
(``full'') polarizabilities.
(b) Spectrum of heat power at distance 200 nm. 
(c) Spectrum at distance $1\,\mu{\rm m}$, same particle size.
}
\label{refr}
\end{figure}

In Fig.~\ref{refr}, we plot the heat exchanged between a SiC particle (at 300 K) 
and a SiC substrate ($T_s = 0$) as a function of the distance. We also show the spectrum of the exchanged flux. We compare, in each plot, Eq.(\ref{eq:spectrum-cooling}) 
to the results of Ref.~\cite{Mulet:2001kp}. We observe that both expressions give 
similar results until at distances as low as 400 nm,
the corrections in the dressed polarizability set in.
The heat spectrum is peaked at a the particle plasmon resonance in the SiC particle, 
where $\epsilon( \omega ) = -2$, i.e., around $935\,{\rm cm}^{-1}$ [Fig.~\ref{refr}(c)].
When one enters in the near field [Fig.~\ref{refr}(b)], two phenomena occur: evanescent contributions to the EM LDOS become stronger, in particular those of the surface phonon polariton resonance where $\epsilon( \omega ) =-1$, around $948\,{\rm cm}^{-1}$,
leading to a double-peaked spectrum.
In addition, the corrections from the dressed polarizability shift the peaks in
relative weight and position. Depending on the distance, these shifts can both enhance 
or reduce the heat transfer compared to the bare polarizability used in Ref.\cite{Mulet:2001kp}, see Fig.~\ref{refr}(a). (The dipolar model is no longer valid at 
the shortest distances shown, however.)
For magnetic particles, similar corrections should be applied.

\section{Conclusion}

We have shown how electromagnetic near fields in the infrared are modified by
the interaction between a probing SNOM tip and the sample it is scanning. 
The tip's response (electric and magnetic polarizabilities) can be modified to account for this interaction. We have combined the effect of retardation to the one of the image dipole. We have analyzed the relation between the signal detected in the far field by an apertureless SNOM and the near fields, excited either coherently (surface plasmon resonance) or incoherently (thermal emission). The technique performs a local spectroscopy of the surface and for some cases the signal can even be proportional to the EM LDOS, a fundamental quantity. 
By comparing to experimental data, we demonstrated one salient feature of strong 
tip-sample interactions, namely interferences between direct and reflected rays that 
lead to oscillatory signals (standing waves) as a function of distance.
We have also corrected the formula for near-field radiative heat transfer in cases where there is a strong particle-sample interaction. In the future, this theory could be extended to tips with more elongated shapes \cite{Huth:2010fg} such as conical ones, often used in experiment, in order to provide a more quantitative comparison. 


\begin{acknowledgments}
The authors thank J.-J. Greffet for fruitful discussions. 
Y. De Wilde and A. Babuty thank the team of R. Colombelli and A. Bousseksou 
for fruitful collaborations on plasmonics. 
Y. De Wilde and A. Babuty acknowledge support from the 
Agence Nationale de la Recherche (Grant No. ANR-07-NANO-039 ``NanoFtir'') 
and from the R\'egion Ile-de-France in the framework of C'Nano IdF 
(nano-science competence center of Paris Region). 
K. Joulain and P. Ben-Abdallah acknowledge support from 
Agence Nationale de la Recherche (Grant No. ANR-10-BLAN-0928-01 ``Sources-TPV''). 
This work is supported by LABEX WIFI  and LABEX INTERACTIFS (Laboratory of
Excellence within the French Program ``Investments for the
Future'') under reference ANR-10-IDEX-0001-02 PSL* and ANR-11-LABX-0017-01.
C. Henkel enjoyed the hospitality of Universit\'e de Poitiers where part
of this work was done.
\end{acknowledgments}


\bibliographystyle{vancouver}

\end{document}